\documentclass[a4,aps,pre,twocolumn,nofootinbib,10pt]{revtex4-1}
\usepackage[latin1]{inputenc}
\usepackage{amsmath}
\usepackage{amsfonts}
\usepackage{amssymb}
\usepackage{graphicx,hyperref}
\usepackage{amssymb}
\usepackage{epstopdf,booktabs}
\usepackage{mathptmx}
\usepackage[usenames,dvipsnames]{color}
\graphicspath{{Figure/}}

\newcommand{\be}{\begin {equation}}
\newcommand{\ee}{\end {equation}}
\newcommand{\beqa}{\begin {eqnarray}}
\newcommand{\eeqa}{\end {eqnarray}}
\newcommand{\mb}{\mathbf}
\hypersetup{
    colorlinks,    citecolor=black,    filecolor=black,    linkcolor=black,    urlcolor=black
}

\begin{document}
\title{Focussing effects in laser-electron Thomson scattering}

\author{Chris Harvey}
\email{cnharvey@physics.org}
\affiliation{Department of Physics, Chalmers University of Technology, SE-41296 Gothenburg, Sweden}
\author{Mattias Marklund}
\email[]{mattias.marklund@chalmers.se}
\affiliation{Department of Physics, Chalmers University of Technology, SE-41296 Gothenburg, Sweden}
 \author{Amol R. Holkundkar}
\email{amol.holkundkar@pilani.bits-pilani.ac.in}
\affiliation{Department of Physics, Birla Institute of Technology and Science - Pilani, Rajasthan, 333031, India}

\begin{abstract}
We study the effects of laser pulse focussing on the spectral properties of Thomson scattered radiation. Modelling the laser as a paraxial beam we find that, in all but the most extreme cases of focussing, the temporal envelope has a much bigger effect on the spectrum than the focussing itself. For the case of ultra-short pulses, where the paraxial model is no longer valid, we adopt a sub-cycle vector beam description of the field. It is found that the emission harmonics are blue shifted and broaden out in frequency space as the pulse becomes shorter. Additionally the carrier envelope phase becomes important, resulting in an angular asymmetry in the spectrum. We then use the same model to study the effects of focussing beyond the limit where the paraxial expansion is valid. It is found that fields focussed to sub-wavelength spot sizes produce spectra that are qualitatively similar to those from sub-cycle pulses due to the shortening of the pulse with focussing. Finally, we study high-intensity fields and find that, in general, the focussing makes negligible difference to the spectra in the regime of radiation reaction.
\end{abstract}

\maketitle

\section{Introduction}
There is currently a great deal of interest in the development of compact, tunable and well-collimated radiation sources. 
Such sources have applications in a wide range of areas including X-ray radiography \cite{Thomlinson:2005}, medical and biological imaging \cite{Neutze:2000}, and in the study of ultra-fast molecular processes.
A method of radiation source generation which is beginning to establish itself is that of nonlinear Thomson/Compton scattering of electrons in intense laser pulses.
Laser-electron setups are much more compact than traditional alternatives such as undulator magnets and magnetic synchrotron rings, thus widening the range of potential applications.
They also show a great deal of promise due to the consistent, exponential increase in peak focal intensities over the past 30 years \cite{Mourou:2006}. With the development of a number of new facilities such as the Vulcan 20 PW upgrade \cite{Vulcan} and the Extreme Light Infrastructure (ELI) Facility \cite{ELI} this trend is expected to continue into the foreseeable future.

The frequency and brilliance of the radiation emitted via Thomson scattering can be adjusted by changing the laser intensity and/or the incoming energy of the electron beam.  At lower intensities one has sufficient control over the laser pulse parameters to generate extremely mono-energetic radiation of a specified frequency. It is also possible to generate frequency combs by manipulating the harmonic structure of the emitted radiation (see, for example, \cite{Heinzl:2010,Harvey:2012MS,Harvey:2009}).  At higher intensities one can generate extremely high-energy, high-brilliance radiation, as has recently been demonstrated in a number of experiments \cite{2011NatPh...7..867C, PhysRevLett.113.224801, doi:10.1117/12.2182569, Downer:16}. Such sources further the range of applications to include both fundamental research \cite{Wu} and more practical applications, such as cancer radiotherapy \cite{Lawrence} and the radiography of dense objects \cite{PhysRevLett.94.025003}.

Many of the theoretical discussions of Thomson/Compton scattering approximate the laser pulse using a plane wave model. In reality the laser pulse will be a focussed electromagnetic field with a more complicated spatio-temporal structure. This is particularly true in the case of high intensity lasers where strong focussing is an important aspect in raising the pulse intensity. 
While there have been a number of works on individual aspects of focussing effects in classical Thomson scattering (see, for example, Ref.~\cite{Lan:2006,Lee:2010}), and in quasi-classical Compton scattering (e.g.~Refs.~\cite{Thomas:2012,Li:2014, Vranic:2015}), the time is ripe for a thorough study of how the structure of a focussed pulse alters the properties of the emitted radiation.
(We also note some promising techniques for tackling the fully quantum case, see Ref.~\cite{DiPiazza_foc:2015}.)
In this work we aim to provide this by systematically analyzing the effects of the laser pulse focussing on the Thomson emission spectra.

We begin in Sec.~\ref{Sec:Model} by discussing the modelling and setup.  In Sec.~\ref{Sec:Results} we consider the effects of focussing on the Thomson spectra, looking at both the importance of the electron impact parameter and the pulse duration. Then in Sec.~\ref{Sec:High-Intensity} we consider high intensity laser pulses, examining the interplay between the pulse focussing and radiation reaction effects. Finally we conclude in Sec.~\ref{Sec:Conclusions}.

\section{Modelling}\label{Sec:Model}
Throughout this work we will consider the case of a relativistic electron in a head on collision with an intense laser pulse. We begin by adopting units such that $c=1$. In all cases the laser will propagate along the $z$ axis, such that its wave vector is $k=\omega_0$, where $\omega_0$ is the central frequency of the laser. We define the peak intensity in the usual manner via the dimensionless parameter $a_0=eE_0/\omega_0 m $.  Unless we are using the vector beam model which we introduce in Sec.~\ref{Sec:Ultra-short}, we will model our laser field as a focussed paraxial beam, the derivation and full expression of which are given in the Appendix \ref{Sec:Paraxial}. Such beams are focussed down to a waist $w_0$ in the centre. The paraxial description itself is a perturbative expansion, satisfying Maxwell's equations to the order of an expansion parameter $\theta_0=w_0/z_r=\lambda/\pi w_0$, where $\lambda$ is the laser wavelength and $z_r=kw_0^2/2$ the Rayleigh length. (In this work we retain terms to fifth order in the expansion parameter.) We denote this quantity by $\theta_0$ since it closely approximates the beam diffraction angle. We consider it a suitable parameter with which to quantify the degree of focussing in our pulses. In all cases our field components are multiplied by a temporal envelope shaping function $a(\eta)$, where the phase $\eta=\omega_0 t-kz$. Except where otherwise stated, we take the field profile to be a Gaussian function
\begin{equation}
a(\eta)=\exp \bigg( -\frac{\eta^2}{2T^2}\bigg ),
\end{equation}
which has a FWHM of $2\sqrt{2\, \textrm{ln} (2)}T$. We note that while this function doesn't strictly satisfy the constraint (\ref{gconstraint}) for all $\eta$, any discrepancy will be in the tails of the pulse where the amplitude will be heavily damped. For the purposes of this study we shall fix the peak intensity as we vary the focussing. This will allow us to better understand the effects of focussing on the locations of structural features in the emission spectrum, but it also means that the total energy content of the pulse will not be conserved with focussing.

\begin{figure}[t]
\includegraphics[width=1.0\columnwidth]{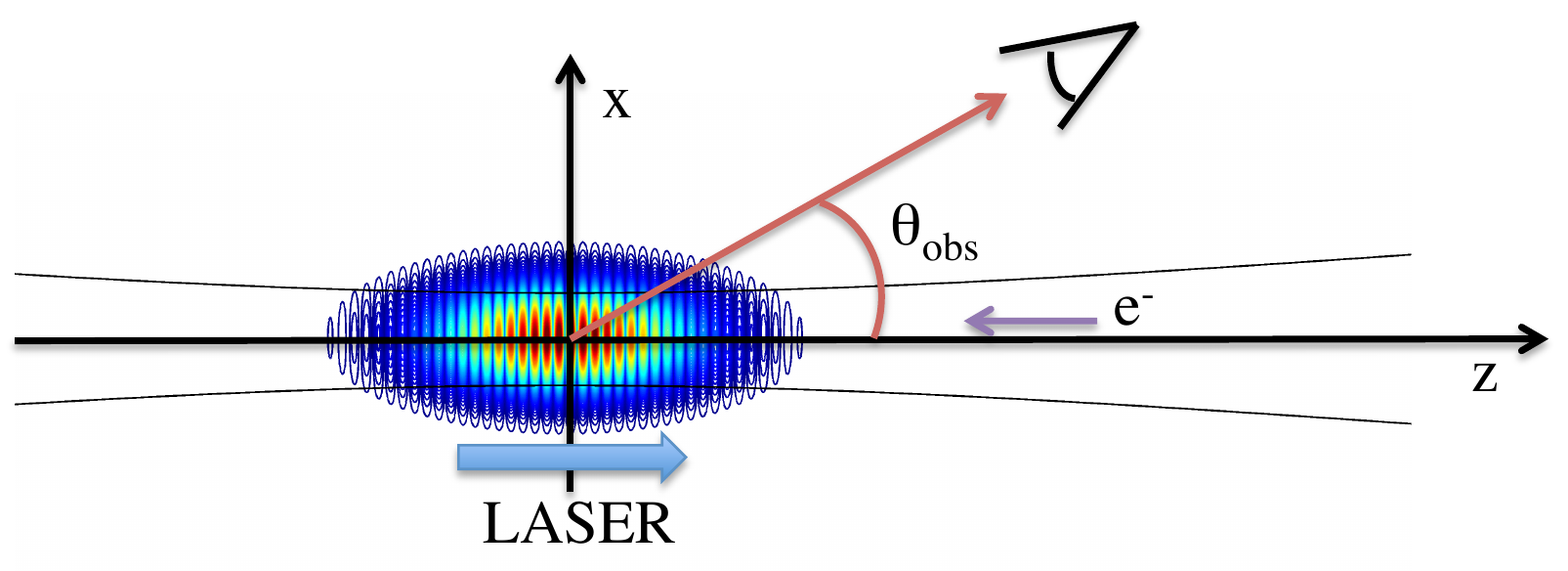}
\caption{Diagram illustrating the coordinate system. \label{fig:setup} }
\end{figure}

\begin{figure}[t]
\includegraphics[width=1.0\columnwidth]{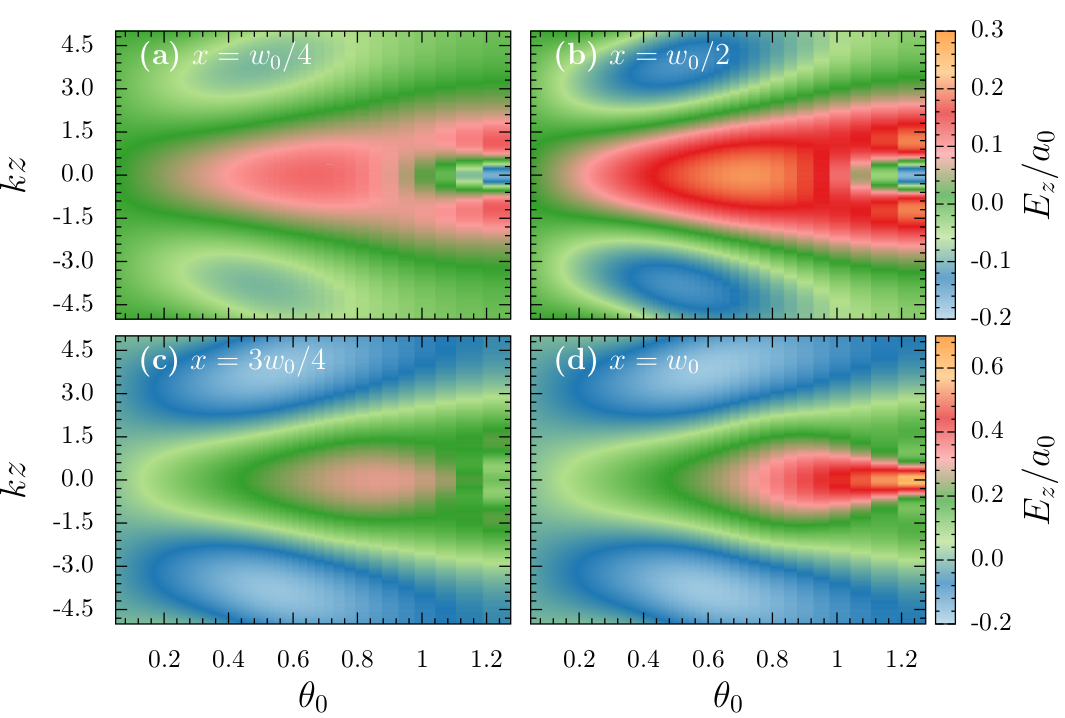}
\caption{Plot showing the (normalised) amplitude of the longitudinal electric field, $E_z(x,0,z)/a_0$, as a function of the focussing parameter $\theta_0$ and the longitudinal coordinate $z$ for different values of $x$. The field component is calculated at $x = w_0/4$ (a), $x = w_0/2$ (b), $x = 3w_0/4$ (c) and $x = w_0$ (d). (It should be noted that as we change the focussing parameter the value of $w_0$ also changes.)  It can be seen that for weak focussing ($\theta_0\rightarrow 0$) the longitudinal field tends to zero.  As we increase the focussing the amplitude rises, becoming a significant fraction of the transverse field. \label{fig:paraxial_eps_vs_z} }
\end{figure}

To conduct our simulations we propagate the electrons through the laser pulses by solving the Lorentz force equation 
\begin{equation}
\frac{d\mathbf{p}}{dt}=e(\mathbf{E}+\mathbf{v}\times\mathbf{B}),
\end{equation}
where $\mathbf{p}=\gamma \mathbf{v}$ is the relativistic momentum and $\mathbf{v}$ the velocity.
Covariance is maintained by enforcing the mass-shell condition $p^2=m^2$ when calculating $\gamma$ (for an alternative method see Ref.~\cite{Harvey:2011}).
Once we have calculated the electron trajectory, the resulting radiation emissions are determined via the well-known classical formula.
The intensity of radiated energy per unit solid angle per unit frequency is given by \cite{jackson},
 \be 
\frac{d^2I}{d\omega\ d\Omega} 
 = \left|\int\limits_{-\infty}^{\infty}\frac{\mb{n}\times[(\mb{n}-{\beta})\times\dot{\mb{\beta}}]}{(1 - \beta\cdot \mb{n})^2}       
 e^{i\ \frac{\omega}{\omega_0} [t + D(t)]} dt \right|^2, 
\label{eq:spectrum}
\ee
where $\mb{n}$ is a unit vector pointing from the particle's position to a detector ($D$) located far away from the interaction, and $\beta$ and $\dot{\beta}$ are, respectively, the particle's {\it relativistic} velocity  and acceleration.  We have normalized the intensity by a factor $e^2/(4\pi^2)$. All the quantities in the expression are evaluated at a retarded time so one can directly do the integration in some finite limit. We illustrate the coordinate system in Fig.~\ref{fig:setup}.
The simulations in this manuscript were carried out using both the code described in Refs.~\cite{Holkundkar:2014,Holkundkar:2015} and the code SIMLA \cite{Green:2015}.

\section{Results}\label{Sec:Results}

\subsection{Effect of focussing}\label{sec:focussing}

We will begin our investigations by restricting ourselves to the case of a counterpropagating electron colliding along the propagation axis of the laser, so that it goes through the centre of the pulse focus. In subsequent sections we will consider the off-axis case.

In the case of a plane wave field, provided the temporal envelope is of sufficient length (and the pulse is symmetric), the net energy change will be zero once the particle leaves the pulse. This is because the acceleration of the particle in the rising part of the cycle is cancelled out by the deceleration in the corresponding down-cycle, an effect resulting from the so-called Lawson-Woodward theorem \cite{Lawson:1946, Lawson:1948}.

\begin{figure}
\includegraphics[width=1\columnwidth]{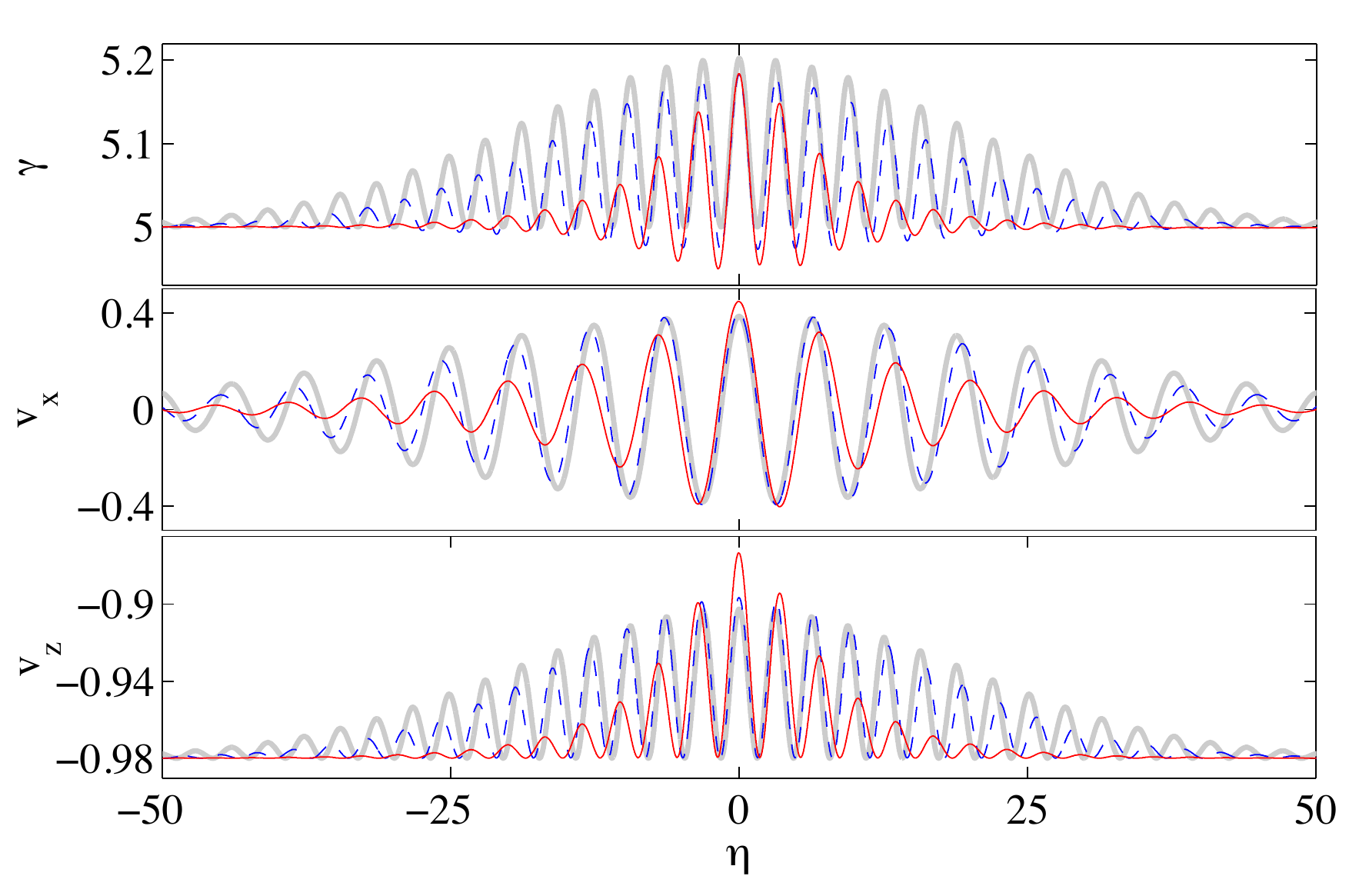}
\caption{Plots showing the $\gamma$ factors and velocity components of the electron in a laser pulse with different focussing. The laser pulse has a Gaussian time envelope and was 27 fs in duration with a peak $a_0=2$. Gray (thick) lines: $\theta_0=0$. Blue (dashed) lines: $\theta_0=0.35$. Red (solid) lines: $\theta_0=0.7$. \label{fig:velocities} }
\end{figure}

\begin{figure}
\includegraphics[width=1\columnwidth]{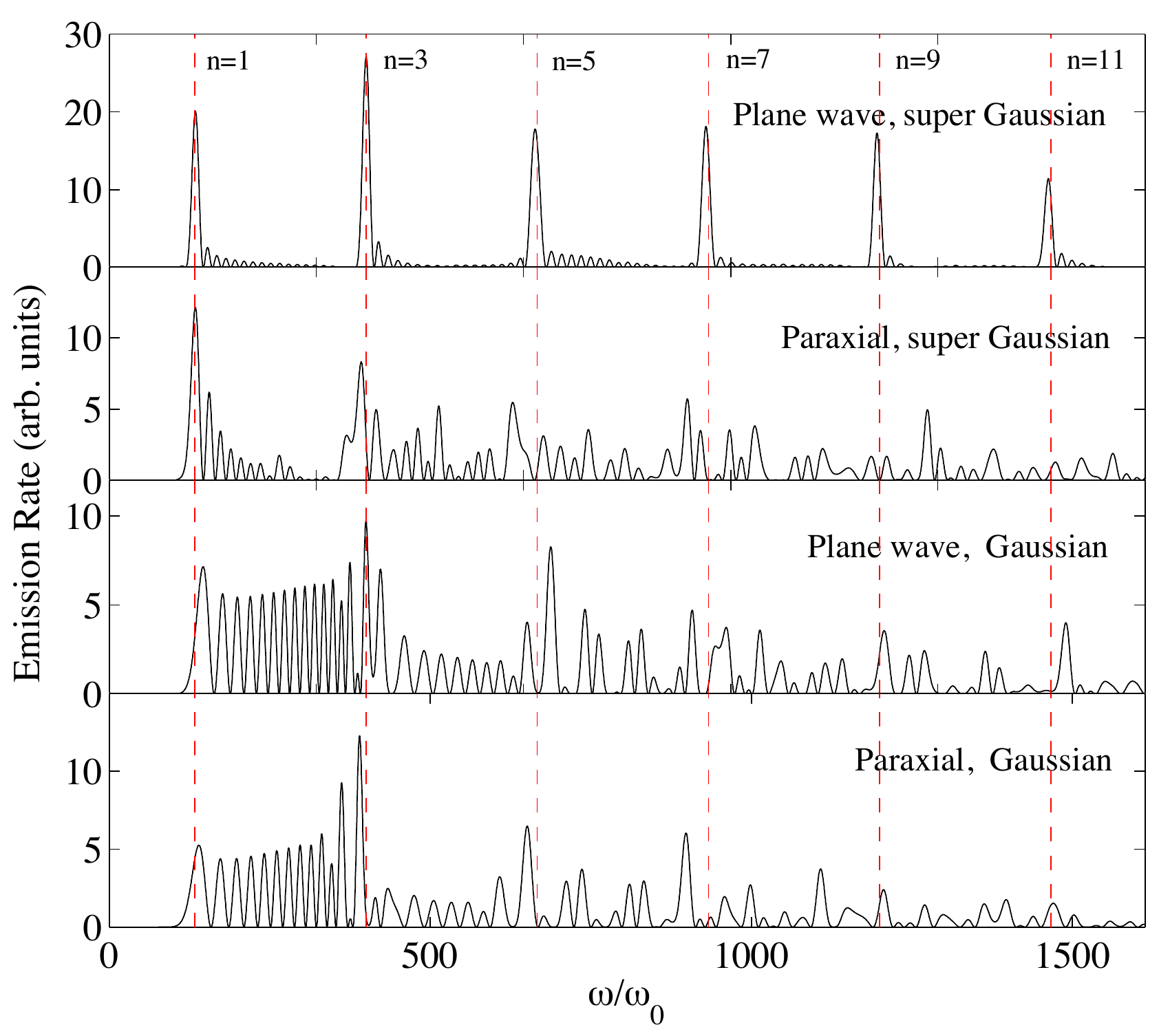}
\caption{Thomson emission spectra calculated at $\theta_{\textrm{obs}}=\pi$ for an electron of initial $\gamma_0=10$ counter-propagating along the laser axis. In all cases the laser pulse has a peak $a_0=2$ and is 10 cycles in duration (27fs). It is either a plane wave or paraxial field (focussed to $\theta_0=1/\pi$ ($w_0=\lambda$)), with a Gaussian or super Gaussian envelope, as indicated in the panels. The red (dashed) lines show the positions of the odd-numbered harmonics calculated for the case of a monochromatic plane wave. \label{fig:spectra1}}
\end{figure} 

As we move from a plane wave to a focussed pulse the field structure will change in a number of ways. The most prominent effects are the development of a longitudinal electric field and a shortening of the pulse duration. The first can be seen from Eq.~(\ref{fieldEz}), where we find that the longitudinal field scales linearly with $\theta_0$, although it is always zero along the laser axis (i.e.~when $x=0$). Nevertheless, the electron will orbit around the axis with the rise and fall of the laser pulse, meaning that it will be subjected to this field even though it is injected along $x=0$. The longitudinal field at different transverse positions is shown in Fig.~\ref{fig:paraxial_eps_vs_z}. The second effect can be seen in Eq.~(\ref{eq:P}) where we see that, in addition to the effect of the temporal envelope function $a(\eta)$, the paraxial beam also decays in longitudinal space like $w_0/w$. Both of these effects become manifest in the electron's velocity which we plot in Fig.~\ref{fig:velocities}, although it is difficult to fully disentangle the two.
We see, for instance, that the tighter the focussing the shorter the interaction time of the electron with the pulse. This is due to the faster fall-off of the field profile, but partially counteracting this effect will be the longitudinal electric field pushing back on the particle. It can be seen from the lower panel of Fig.~\ref{fig:velocities} that the longitudinal velocity decreases in magnitude (note the scale is negative) in the centre of the tightly focussed pulse, which results in a reduction in the $\gamma$ factor (top panel), but has slightly lesser impact on the transverse velocity (centre panel).

We now turn our attention to the Thomson emission spectra. In the nonlinear case where $a_0>1$ the total emission spectrum will be composed of a sum of harmonic contributions, each corresponding to integer multiples of the laser frequency \cite{Sarachik:1970, Esarey:1993}. In the idealised case of an infinite monochromatic plane wave field the harmonics will be very narrow, tending to $\delta$ function spikes as $a_0\rightarrow\sqrt{2}\gamma$ \cite{Harvey:2009}. The positions of these spikes in frequency space are determined by considering the conservation of the cycle averaged momentum \cite{Harvey:2012MS}, and for the case of a relativistic particle, $\gamma\gg 1$, if observed in the backscattering direction,  $\theta_{\textrm{obs}}=\pi$, they are \cite{Harvey:2009}
\begin{equation}
\omega_n^\prime = \frac{4\gamma^2 n\omega_0}{1+\frac{a_0^2}{2}+4\gamma n\frac{\omega_0}{m}},\label{eq:omega_n}
\end{equation}
where $n$ is an integer corresponding to the harmonic number. 
(From the cycle-averaged momentum $q$ one can define a cycle averaged {\it effective} mass $m_{\ast}^2 =q^2 = m^2 (1+a_0^2/2)$. It is found that the frequency spectrum behaves as if the electron mass had become ``dressed'' by the background field, e.g.~the positions of the harmonics are shifted by a factor $1+a_0^2/2$ \cite{Sengupta:1952,Kibble:1966,Harvey:2009,Harvey:2012MS}.)
In the case of linear polarisation only the odd numbered harmonics will contribute to the spectrum in the on-axis direction ($\theta_{\textrm{obs}}=\pi$) \cite{Esarey:1993}. 
Although in reality it's not possible to generate an infinite plane wave field, we can approximate one quite closely be defining our temporal envelope to have a super Gaussian profile \cite{Harvey:2012MS, Harvey:2015NUM} 
\begin{equation}
a(\eta)=\exp \bigg( -\frac{\eta^{12}}{2T^{12}} \bigg ).
\end{equation} 

\begin{figure*}[ht]
\begin{center}
\includegraphics[width=0.8\textwidth]{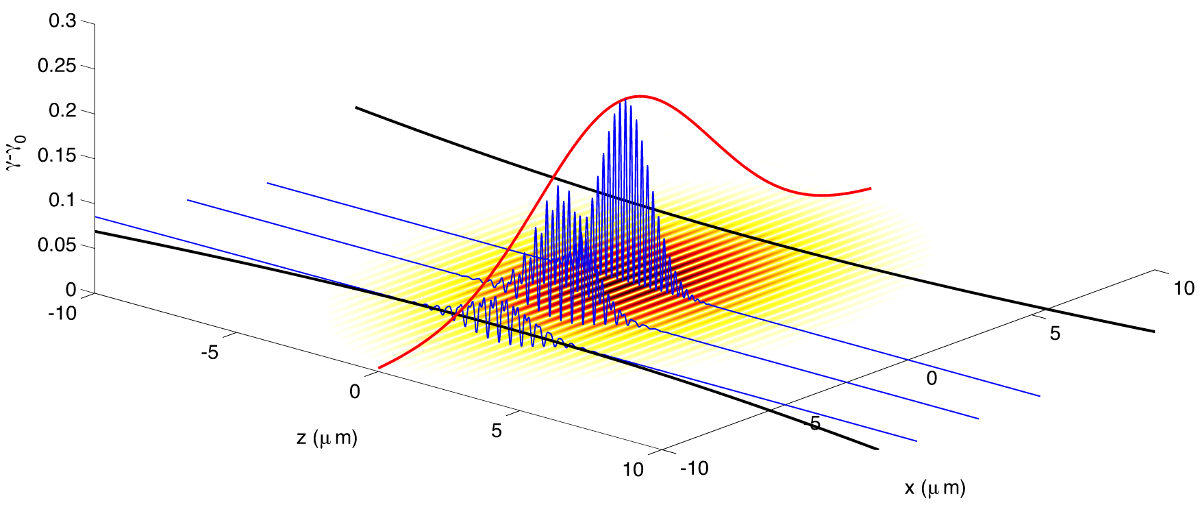}
\caption{Illustration showing the electron $\gamma$ factor as it propagates through a focussed pulse with varying impact parameter. In this example the electron had an initial $\gamma_0=5$ and was inserted into the laser with impact parameters of $x_i=0$, $x_i=w_0/2$ and $x_i=w_0$, which correspond to the three blue lines in the plot (back to front, respectively).  The laser pulse had a peak $a_0=2$, $\lambda=0.8\, \mu m$ and was focussed to a waist of $w_0=5\, \mu m$ ($\theta_0=0.05$). The duration was reduced to 20 fs to aid clarity of illustration. The yellow-black shading illustrates the intensity profile at $t=0$. The curved black lines in the $x$-$z$ plane show the beam waist $w$. The red curve in the $x$-$\gamma$ plane shows the transverse profile function, $\exp(-r^2/w^2)$, evaluated at $y=z=0$. \label{fig:impact_parameter} }
\end{center}
\end{figure*}

It is well known that the introduction of a more smoothly decaying temporal profile, such as a Gaussian, results in a broadening out of the harmonics over a finite frequency interval \cite{Heinzl:2010, Boca:2009}. This is because there are now significant contributions to the emissions from the electron radiating in the rise and fall of the pulse. Thus it makes sense for us to consider four cases: the emissions from an electron in a plane wave and a paraxial field, and with Gaussian and super Gaussian envelopes. Comparing these will allow us to disentangle the changes to the emission spectrum that result from the focussing of the fields from those that result from the shape of the temporal profile. In Fig.~\ref{fig:spectra1} we show the four cases calculated for a laser of peak intensity $a_0=2$, duration 27 fs (10 optical cycles), $\lambda=0.8\, \mu m$ and for an incoming electron with initial $\gamma_0=10$. Inserting these values into Eq.~(\ref{eq:omega_n}) tells us that the separation between harmonics is about 207 eV. We have marked the position of the (monochromatic field) harmonics as dashed red lines on the plot. It can be seen in the top panel that, as expected, the plane wave field with a super Gaussian envelope is a good approximation to the infinite plane wave field, with all the harmonics lining up in the correct positions and being very narrow, almost $\delta$ function, spikes.  If we retain the super Gaussian envelope but switch to a focussed paraxial beam ($w_0=\lambda$, $\theta_0=1/\pi\approx 0.32$), the harmonics begin to spread out across a range of frequencies and as a result there is a reduction in the peak amplitudes (second panel). We now switch back to a plane wave, but this time with a Gaussian time profile. The result (third panel) is a much more dramatic broadening of the harmonics than occurred from focussing the field, even though the focussing we used was quite strong. (Unfortunately, the highly nonlinear relationship between the pulse profile and the properties of the sub-harmonics means that we are unable to quantify these substructures in more detail. However, there are good discussions of their properties in Refs.~\cite{Heinzl:2010, Boca:2009}.) Finally, in the bottom panel we consider a focussed pulse with a Gaussian time envelope. Overall we conclude that, although focussing the field results in a broadening of the harmonics, the effect is significantly smaller than the effect of the temporal profile of the field, be it plane wave or focussed.

\subsection{Effect of impact parameter}\label{Sec:impact}

\begin{figure}
\includegraphics[width=1\columnwidth]{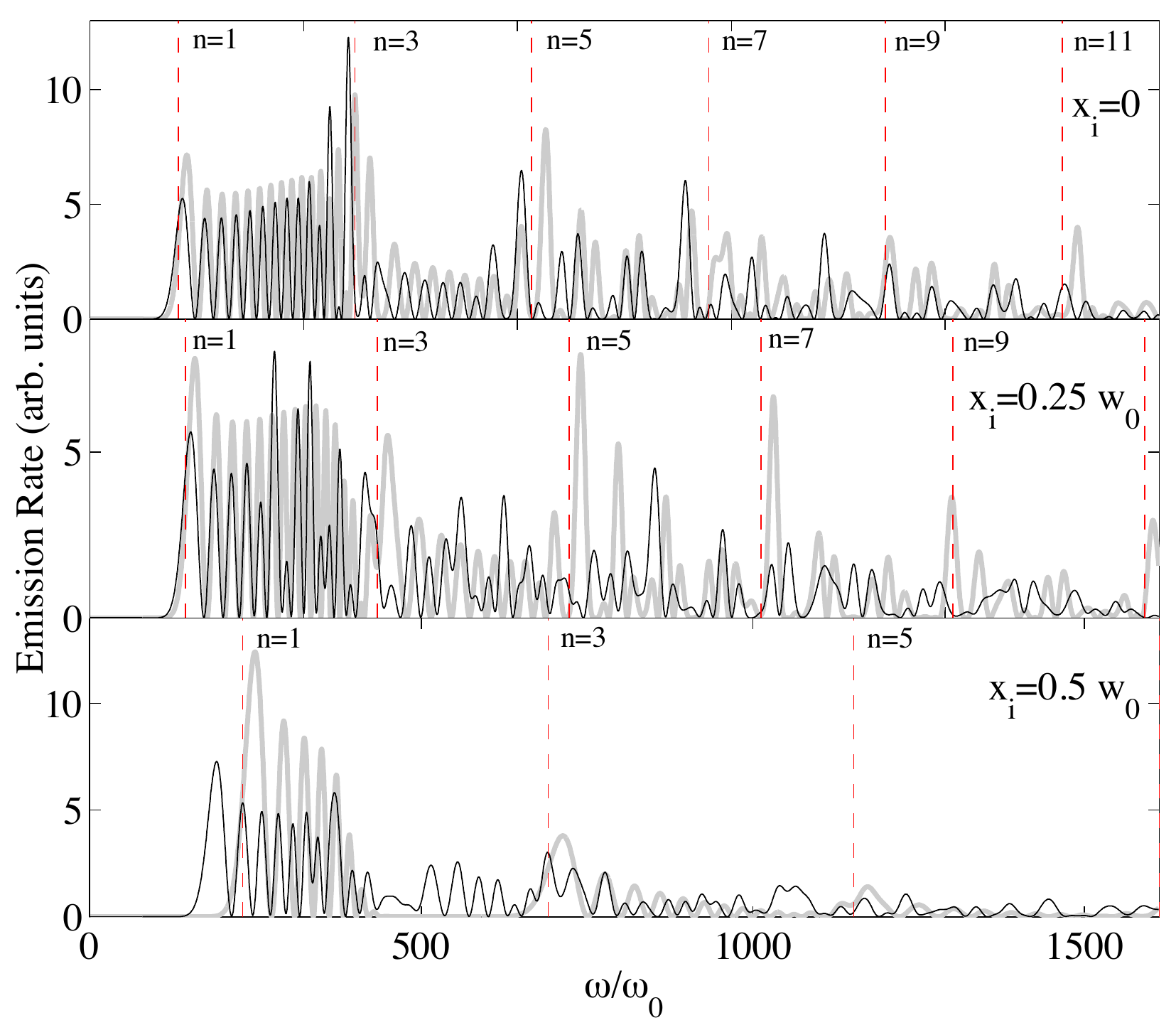}
\caption{Thomson emission spectra for electrons with varying impact parameters $x_i$. The laser is of peak intensity $a_0=2$, $\lambda=0.8\mu m$, and with a Gaussian profile of 27fs duration. The incoming electrons have an initial $\gamma_0=10$ and are counter-propagating with the laser pulse. Their impact parameters are listed in the figure panels. Black lines: emission spectra calculated for a paraxial laser pulse focussed to $\theta_0=1/\pi$, ($w_0=\lambda$). Gray (thick) lines: emission spectra calculated using a plane wave laser, with peak amplitude reduced to compensate for the fall-off in the focussed field (see main text for details). Red (dashed) lines: the positions of the harmonics in the case of a monochromatic field, calculated using the same $a_0$ as the plane wave spectra.\label{fig:impact_spectra}}
\end{figure} 

In a real experiment the electron beam will have a finite transverse width and so we must consider how the emission spectrum changes for particles with a non-zero impact parameter.  It can be anticipated that, to a certain extent, the full emission spectrum from all the electrons can be approximated by the spectrum produced from the on-axis electron.
 This is because the electrons close to the axis will see a stronger field and so their emissions will dominate over the others.  In the case of plane waves the result is trivial: regardless of the impact parameter we will always obtain the same spectrum. We will now investigate for a focussed pulse.

In Fig.~\ref{fig:impact_parameter} we illustrate the evolution of the electron $\gamma$ factor as electrons with different impact parameters pass through a (weakly) focussed pulse. It can be seen that as we increase the impact parameter two main changes occur. The first is that the amplitude of the $\gamma$ factor decreases, as should be expected since the field decays like $\exp(-r^2/w^2)$ (see Eq.~\ref{eq:P}). The second is that the electrons can momentarily have a lower $\gamma$ factor than they started with as they oscillate in the field. The result of this is that the peak $\gamma$ factor is lower than would otherwise be expected, and is not so well approximated by just considering an exponential fall-off of the form $\exp(-r^2/w^2)$, which is plotted in red. Finally, a careful observation finds that the $\gamma$ factors are not symmetric along $z$, resulting in a small net energy change as the electrons exit the pulse.

We now turn our attention to the Thomson spectra, considering once again a laser of peak intensity $a_0=2$, $\lambda=0.8\,\mu m$, and with a Gaussian profile of 27 fs duration. The incoming electrons have an initial $\gamma_0=10$ and are counter-propagating with the laser pulse, although they now have impact parameters in the $x$ (polarisation) direction of $x_i=0$, $w_0/4$ and $w_0/2$. The resulting spectra are shown as black lines in Fig.~\ref{fig:impact_spectra}. It can be seen that both the emission amplitudes and the number of harmonics become increasingly damped as we increase the impact parameter. This is to be expected since the electrons are now probing regions of field with lower peak intensity. Also in Fig.~\ref{fig:impact_spectra} we plot as gray lines the emission spectra for a plane wave with peak amplitude reduced by a factor
\begin{equation}
a_0 \rightarrow a_0\exp\bigg(-\frac{x_i^2}{w^2}\bigg)\Bigg\vert_{y=z=0}=a_0\exp\bigg(-\frac{x_i^2}{w_0^2}\bigg),
\end{equation}
to compensate for the transverse fall-off in intensity for the focussed field. (Additionally we plot the locations of the monochromatic harmonics for this case using red dashed lines.) It can be seen that this modified plane wave approximation is only of limited value in approximating the emission spectra for large impact parameter. It predicts the fall-off with frequency with reasonable accuracy, but the structural features of the harmonics are very different as compared to the focussed field.

\begin{figure}
\includegraphics[width=1\columnwidth]{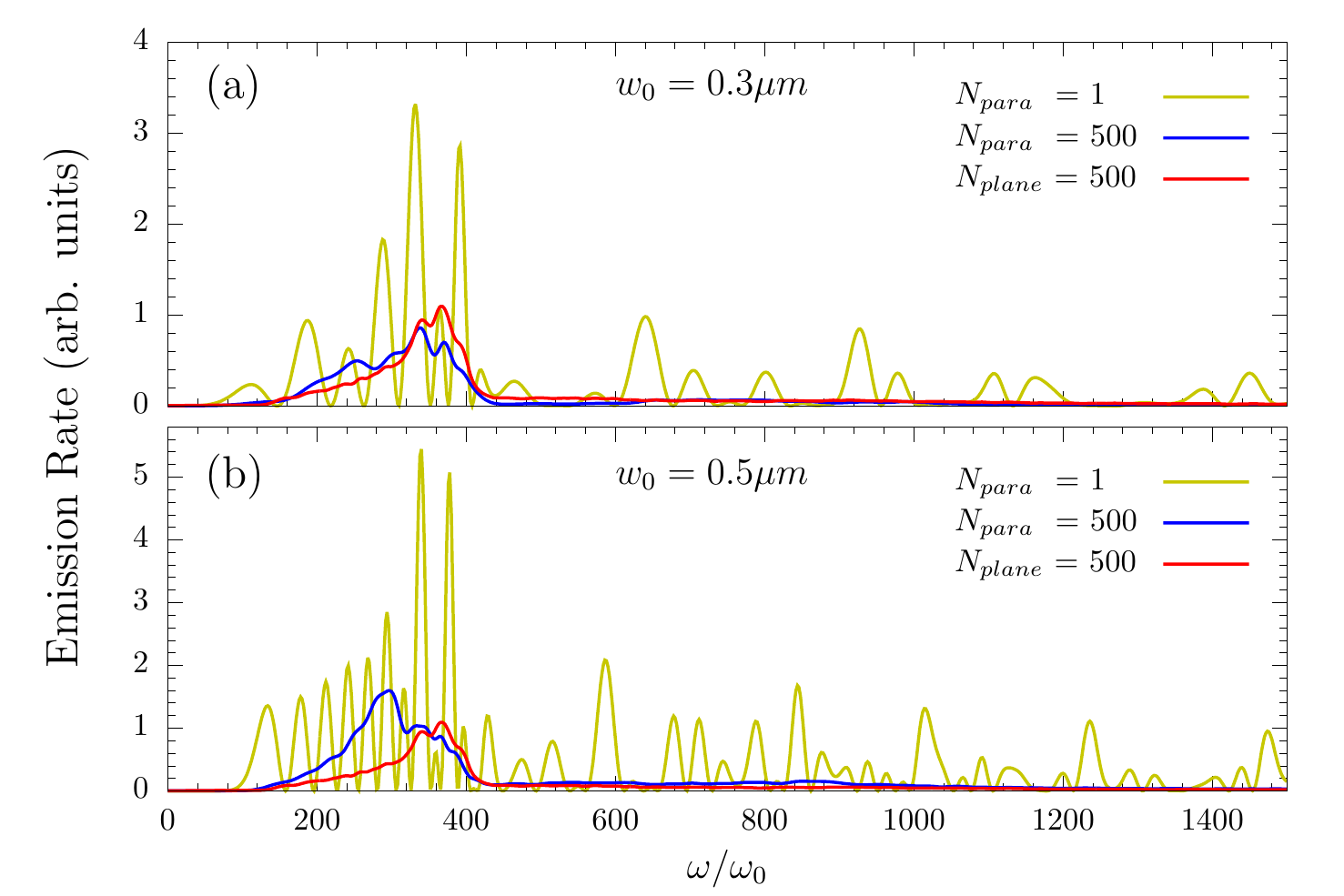}
\caption{Thomson emission spectra for an electron bunch in a paraxial beam (500 particles randomly distributed over a disk of radius $w_0$ in the transverse plane, shown as blue lines and denoted $N_{para}=500$) are compared with those of a single particle (yellow lines, denoted $N_{para}=1$). The electron(s) have initial $\gamma_0 = 10$ and collide head-on with $0.8 \mu m$, 10 cycle laser beam of peak intensity $a_0 = 2$. Panel (a) shows the results for a focussing of $\theta_0=0.85$ ($w_0 = 0.3 \mu m$), and panel (b) for $\theta_0=0.51$ ($w_0 = 0.5 \mu m$).  The spectra is also calculated for the bunch using a plane wave field multiplied by a transverse envelope function $\exp(-r^2/w_0^2)$ (red lines, denoted $N_{plane}=500$). \label{fig:elec_bunch}}
\end{figure}

Since the effect of the impact parameter has most relevance in the context of a bunch of electrons colliding with the laser pulse, we now consider a spatially distributed group of electrons. In order to keep the analysis manageable we will limit ourselves to a two-dimensional disk of electrons in the transverse plane. (If we were to extend the disk to a cylinder we would find that the electrons at the ends of the cylinder will arrive early/late, before the pulse has reached its peak focus or after the focus has decayed. In such instances the emissions from these electrons would smear out the tails of the combined spectrum, making our analysis more difficult.) In Fig.~\ref{fig:elec_bunch} we compare the total (normalised) emission spectrum from a bunch of electrons with that from a single particle travelling along the axis. It can be seen that, in the case of the first harmonic,  the (averaged) emissions from the electron disk (blue lines) are quite well described by the single on-axis electron, forming a spectrum that is approximately an average over the structures of the single particle spectrum. However, the higher harmonics in the spectra are missing in the case of the electron disk. This is because most of the electrons pass through weaker regions of field than exist at centre and for these electrons the higher harmonics are heavily damped.  Thus we conclude that the lower frequency part of the spectrum is reasonably well described by the single electron, but the higher part is not. Finally, we also plot the emission spectrum for the disk of electrons in collision with a plane wave multiplied by a transverse envelope function $\exp (-r^2/w_0^2)$ (red lines). Doing so allows us the distinguish the changes to the spectrum due to the transverse fall off of the focused field from other focussing effects. It can be seen that the modified plane wave spectrum represents the true spectrum (i.e.~the blue line) reasonably well. Although it significantly underestimates the lower part of the main harmonic, it reproduces the correct fall-off of the higher harmonics. (It should be noted that this is not identical to what we have done to produce the gray lines in Fig.~\ref{fig:impact_spectra}; there we reduced the peak intensity by a constant factor, in the current example we are reducing it by a factor that is a function of the transverse coordinates.)

\subsection{Ultra-short pulses}\label{Sec:Ultra-short}
In the case where the pulse duration is very short the paraxial approximation will no longer be valid. This is because in such a situation the expansion parameter $\theta_0$ (which closely approximates the beam diffraction angle) will be of a similar order to the the timescale of the pulse duration $\omega_0 T$, and so the fields no longer vary gradually along the propagation axis. Instead we adopt the vector beam model described in Sec.~\ref{Sec:Vector_Beams}, which is derived from an oscillating dipole field \cite{Lin:2006}. This provides an exact analytical solution to Maxwell's equations describing a focussed field with an arbitrarily short duration.

\begin{figure}
\begin{center}
\includegraphics[width=0.9\columnwidth]{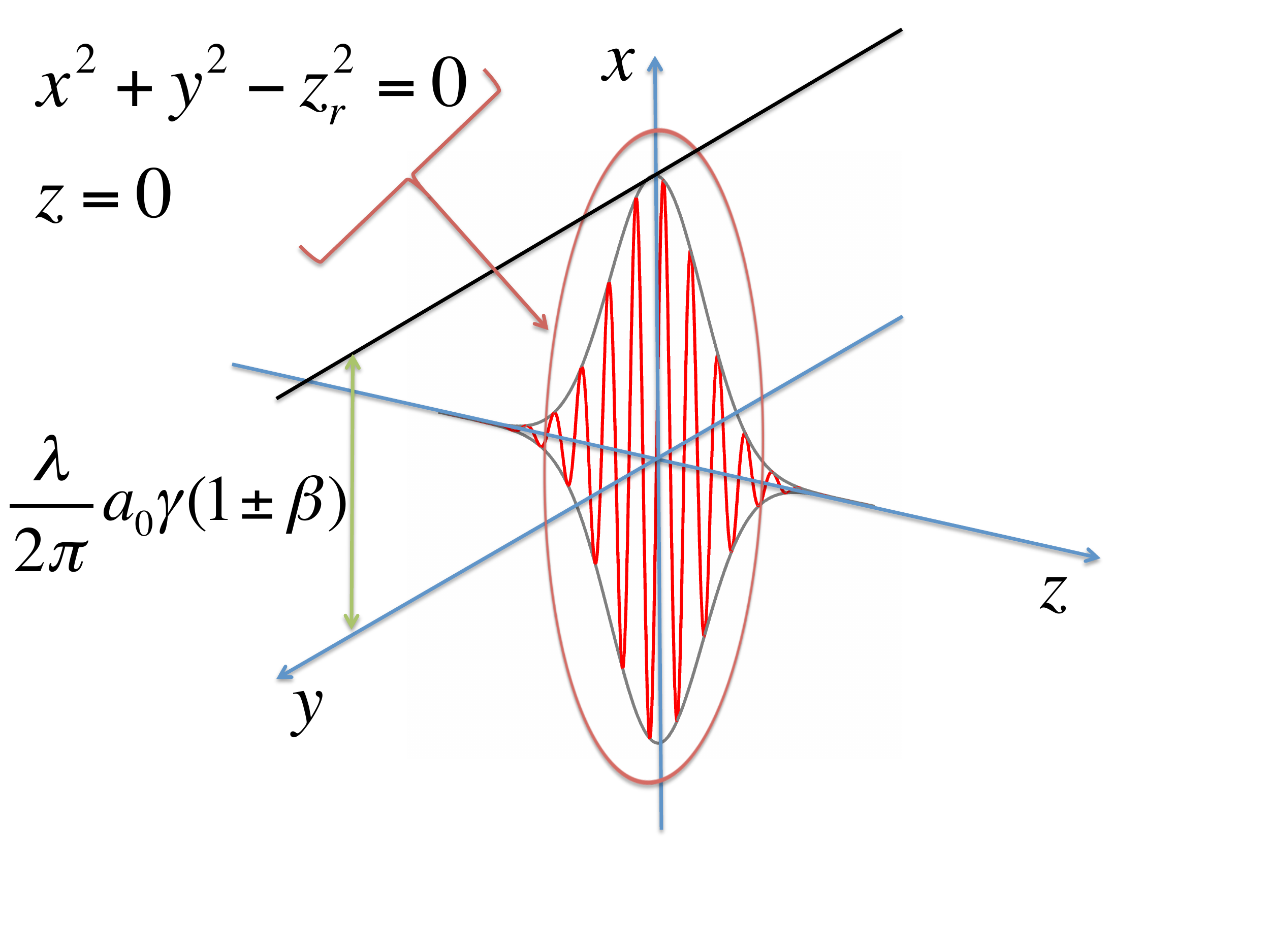}
\caption{Diagram showing the location of the singularity in the vector beam model in relation to the particle orbit. \label{fig:ring_singularity} }
\end{center}
\end{figure}

However, the vector beam model is not without problems of its own. The most notable is the presence of a ring singularity caused by the fields blowing up when the complex distance $R=\sqrt{x^2+y^2+(z+iz_r)^2}\rightarrow 0$ (see Sec.~\ref{Sec:Vector_Beams} for details). A quick calculation shows that this occurs when $x^2+y^2+z^2-z_r^2=0$ and $2izz_r=0$. Thus the fields are not properly defined along a ring, centred at the origin and of radius $z_r$.  In order to assess the impact of this we must consider the motion of the electron in the field. If we take the electron to be on axis with zero impact parameter, then the radius of the orbit can be estimated by considering that of a particle in a plane wave. This can easily be determined analytically (see, e.g., Appendix B of Ref.~\cite{Taub:1948}) and is given by
\begin{equation}
x_{\perp,\textrm{max}}=\frac{\lambda}{2\pi}a_0 \gamma  (1\pm \beta),
\end{equation}
where the sign $\pm$ is positive (negative) for co- (counter-) propagating particles. Thus to be confident that the electron will not come into the vicinity of the singularity we require
\begin{equation}
a_0 \gamma  (1\pm \beta) < \frac{2\pi^2 w_0^2}{\lambda^2}.
\end{equation}
Assuming $w_0\sim\lambda$ we can see that there will only ever be a problem in the counter-propagating case (with zero impact parameter) that we are considering if $a_0\gg \gamma$.  We illustrate this argument in Fig.~\ref{fig:ring_singularity}.

\begin{figure}
\includegraphics[width=1.0\columnwidth]{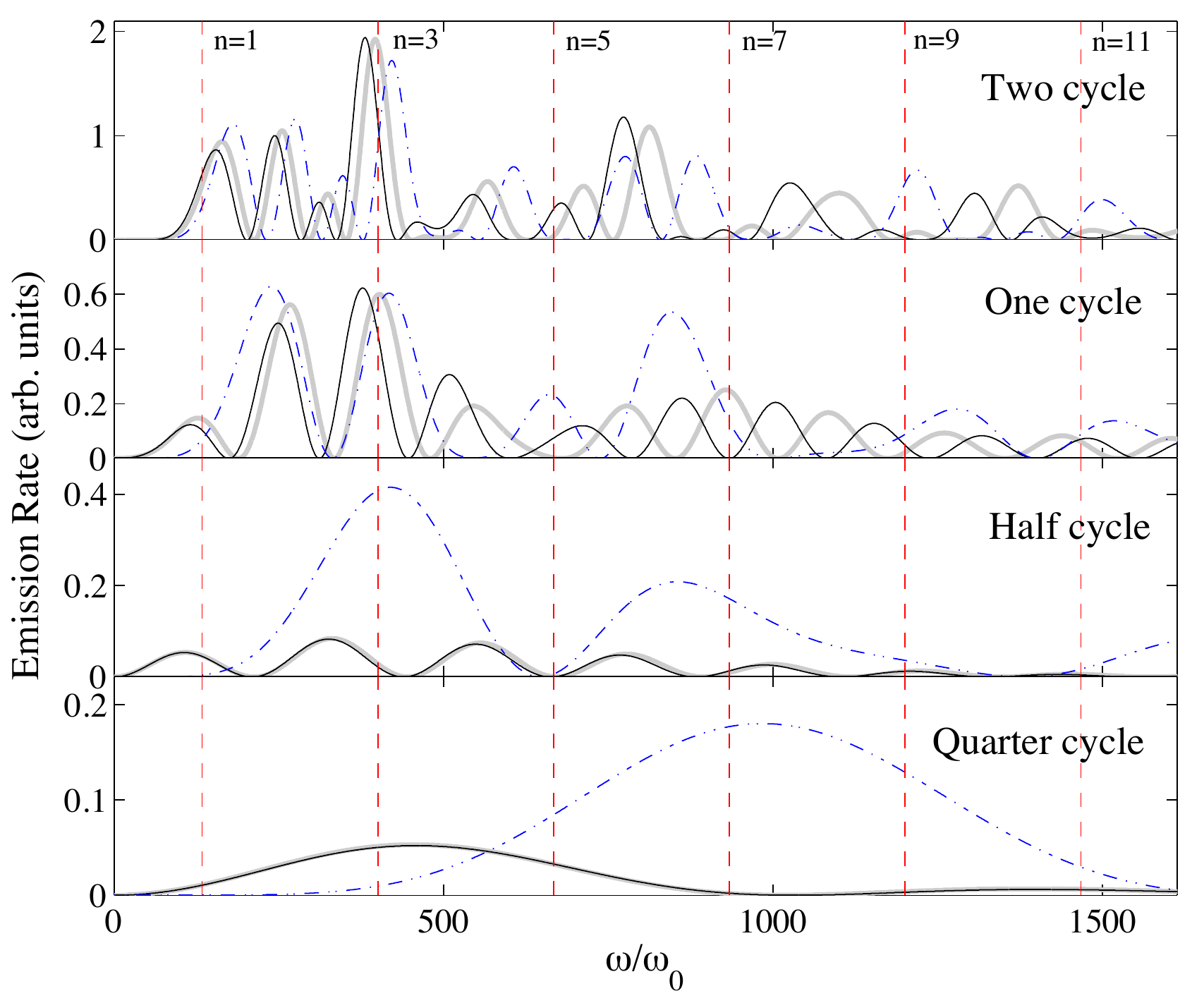}
\caption{Emission spectra for laser pulses of decreasing duration (FWHM). The laser is of peak intensity $a_0=2$ and wavelength $\lambda=0.8\mu m$. The electron has an initial $\gamma_0=10$. Gray (thick) lines: spectra calculated for a plane wave. Black (solid) lines: spectra calculated using a 5th order paraxial beam focussed to a spot size of $w_0=0.8\mu m$ ($\theta_0=1/\pi$). Blue (dash-dot) lines: the same but modelled using the vector beam model. Red (dashed) lines: the locations of the harmonic peaks for the case of an infinite plane wave (see Sec.~\ref{sec:focussing}). \label{fig:ultra-short} }
\end{figure}

To consider the effects of the pulse duration on the Thomson spectra we once again work with a laser pulse of peak intensity $a_0=2$ brought into collision with an electron counter-propagating along the laser axis with $\gamma_0=10$. A series of spectra are plotted in Fig.~\ref{fig:ultra-short}. The top panel shows spectra for pulses containing two optical cycles FWHM and in each subsequent panel the pulse durations are reduced by half. Although the two-cycle spectra are different from the longer pulses shown in Fig.~\ref{fig:spectra1}, the qualitative features are broadly the same. For instance the positions of the spectral peaks are close to the locations predicted by the infinite plane wave model (vertical red lines). As we decrease the pulse duration the harmonics reduce in amplitude and become blue shifted to higher frequencies. Neither of these facts is surprising. The reduction in amplitude roughly scales with the reduction in duration and can thus be interpreted as a result of the electron radiating for a shorter time. (We also note that, in order to simplify our analysis, we have fixed the peak intensity, rather than the total pulse energy, meaning that as we increase the focussing we reduce the total energy content of the pulse.) The frequency blue shift results from the fact that the intensity dependent mass, $m_{\ast}$, only governs the harmonics when the laser field has sufficient periodicity \cite{Harvey:2012MS,PhysRevA.83.032106}. In the case of a sub-cycle pulse the mass is no longer dressed by the laser field and so the harmonics are blue shifted back to what would be expected from the  Klein-Nishina formula for a single photon scattering off an electron (see, e.g., Ref.~\cite{Harvey:2009}). This behaviour is consistent with what is found in the QED case for Compton scattering in sub-cycle pulses \cite{PhysRevA.83.032106} (see also Ref.~\cite{PhysRevA.83.022101}).

\begin{figure}
\includegraphics[width=1.0\columnwidth]{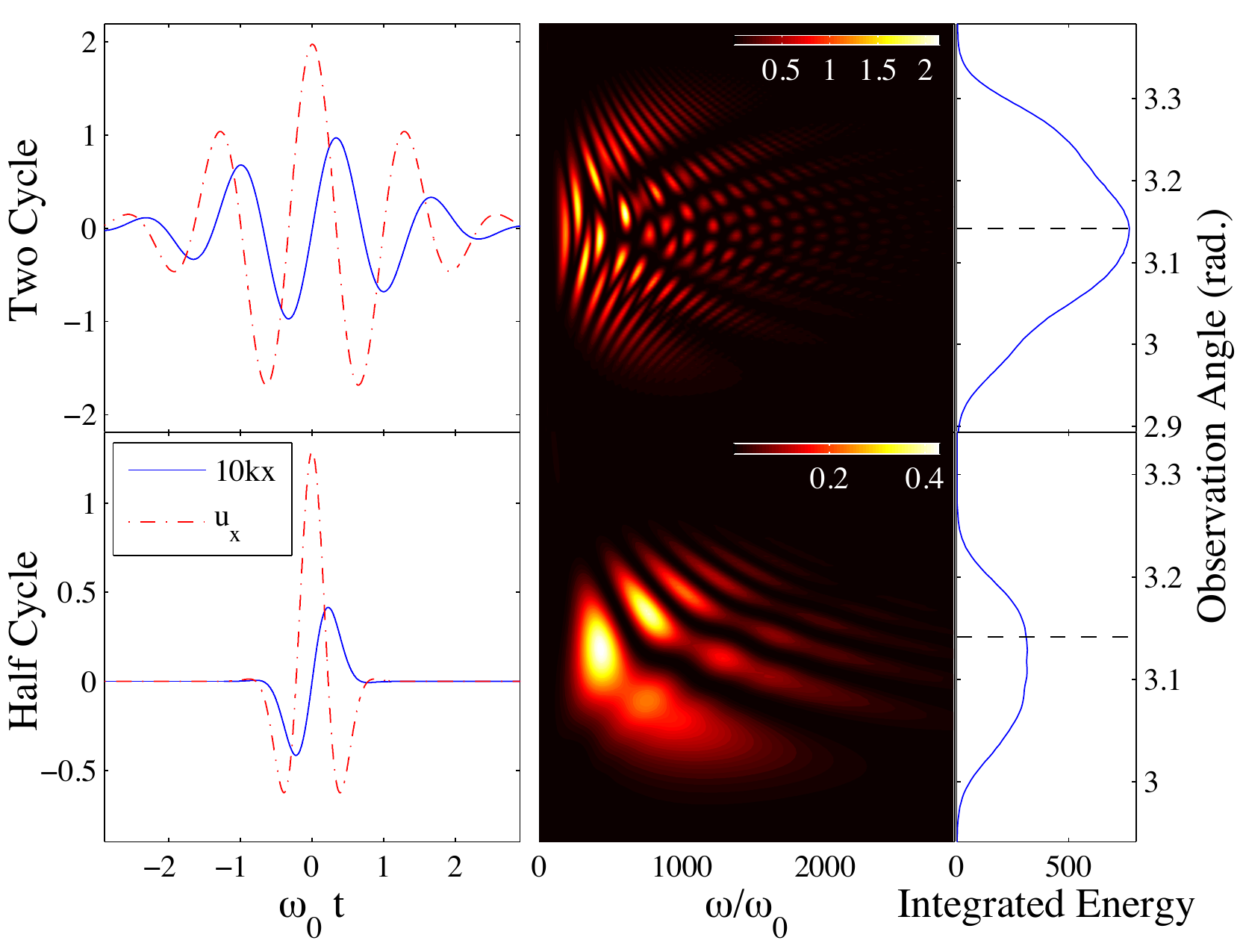}
\caption{Electron motion and corresponding emission spectra in short laser pulses. The laser is described using the vector beam model and has a peak $a_0=2$, $\lambda=0.8\mu m$ and is focussed to a spot size of $w_0=0.8\mu m$ ($\theta_0=1/\pi$). The electron collides along the propagation axis and has an initial $\gamma_0=10$. Top row: results for a 2 cycle (FWHM) pulse. Bottom row: results for a 0.5 cycle (FWHM) pulse. Left column: transverse velocity, $u_x$, and transverse trajectory $10\times k\cdot x$ (units dimensionless). Centre column: emission spectra as a function of frequency and angle. Right column: angular emission spectra integrated over all frequency (arb. units). The dashed line indicates the propagation direction ($\theta=\pi$). (All units dimensionless.)\label{fig:ultra-short2D} }
\end{figure}

In order to disentangle the effects of focussing from the effects of the short pulse duration we have calculated the spectra for the plane wave, paraxial and vector beam pulse models. For the two cycle pulse there is reasonable agreement between all three descriptions. As we shorten the pulse duration the vector beam results begin to separate from the paraxial and plane wave results.  Although the amplitude is damped compared to that of longer cycle pulses, the emission spectra predicted by the vector beam model has a significantly higher amplitude than that predicted by the paraxial/plane wave models. The radiation from the vector beam also has a higher frequency than that predicted by the other models. This is due to the fact that the vector beam description contains a so-called ``self-induced blue shift'' \cite{Lin:2006}. This is analogous to the Gouy phase, which is a frequency shift that occurs as a result of focussing, whereas the self-induced blue shift is caused by the finite pulse duration, becoming important in the sub-cycle regime. The higher-frequency components of the vector beam pulse will increase the frequency of the scattered radiation and raise its amplitude, since higher frequency fields contain more energy. These examples demonstrate how crucial it is to adopt a suitable beam model when considering sub-cycle pulses.

Finally, in Fig.~\ref{fig:ultra-short2D} we show the angular emission spectra for two (vector beam) pulses of different durations. It can be seen that as the pulse duration decreases the angular distribution of the radiation loses its symmetry. This effect is visible even for the two-cycle pulse, but becomes quite noticeable in the case of the half-cycle pulse (the spectrum for a longer pulse can be seen in panel (b) of Fig~\ref{fig:tight2d}). The reason for the breakdown of symmetry can be found by studying the electron motion. From the lefthand panels in Fig.~\ref{fig:ultra-short2D} it can be seen that, close to the centre of the pulse (around $t=0$), the longitudinal velocity, $u_x$, mostly has a positive value only. Since the majority of the emissions will occur while the electron is in this region, and the radiation will be emitted approximately in the direction of motion \cite{jackson}, the spectrum will be skewed towards the positive $x$ direction ($\theta_{\textrm{obs}}<\pi$). This is of course an effect of the carrier envelope phase, with the short duration amplifying the field asymmetry caused by the phase difference between the laser carrier wave and the pulse envelope. We could correct for this by phase shifting the vector field by half a cycle to make it a cosine pulse. However, the asymmetry in the radiation direction offers a useful diagnostic tool for determining the carrier envelope phase of a given laser pulse \cite{Mackenroth:2010, Titov:2016}.

\subsection{Extreme focussing}\label{Sec:Extreme-focussing}

\begin{figure}[t]
\includegraphics[width=1\columnwidth]{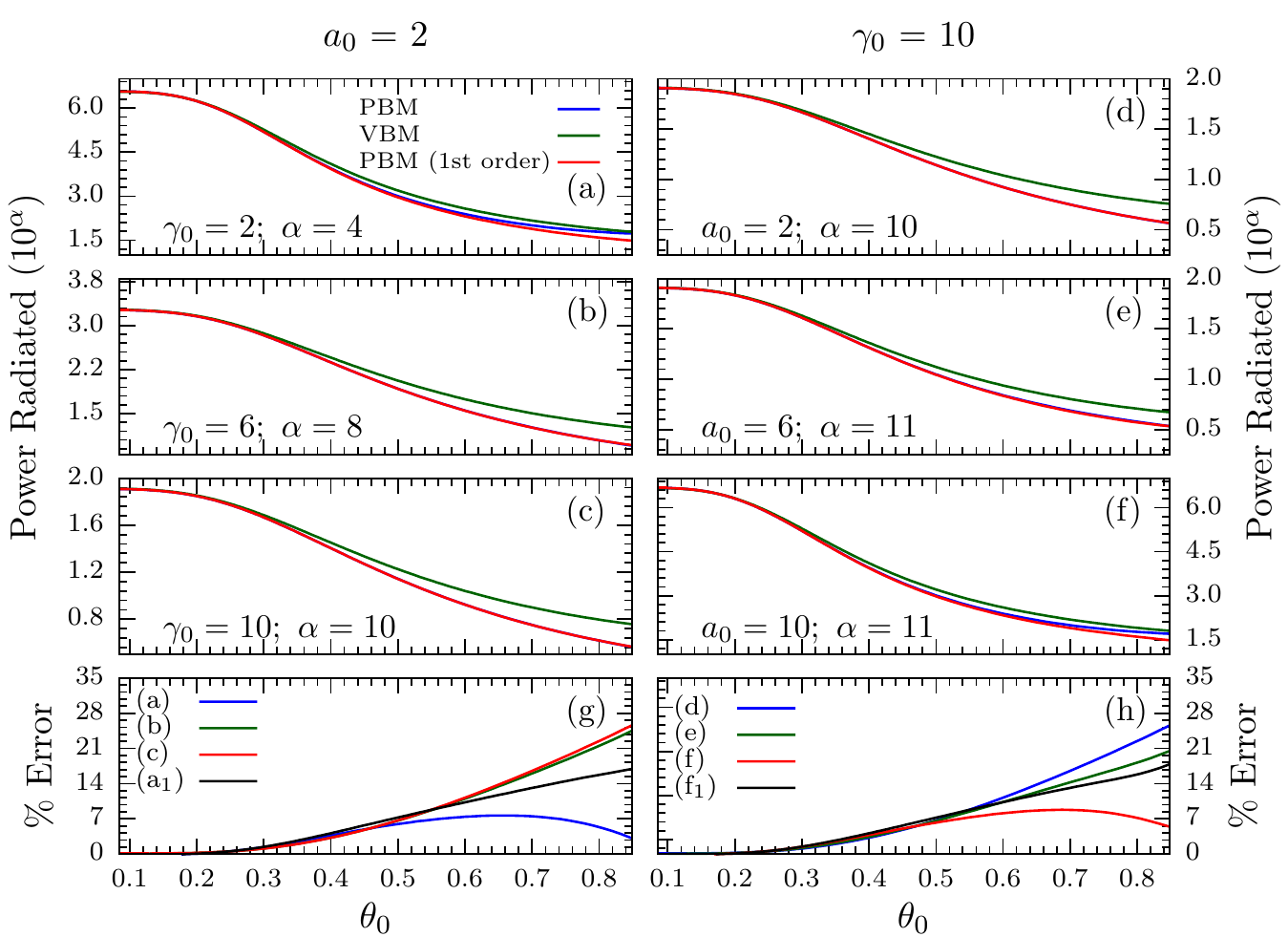}
\caption{Plots showing the total radiated power (Larmor power) in dimensionless units for fixed $a_0$ (a,b,c) and for fixed $\gamma_0$ (d,e,f) using the Vector Beam Model (VBM) and the 1$^{st}$ and 5$^{th}$ order Paraxial Beam Model (PBM). The relative $\%$ error in radiated power for the paraxial beam with respect to the vector beam model is presented in (g) and (h). The lines denoted by (a$_1$) and (f$_1$) are for the case of the 1$^{st}$ order paraxial beam, in all other cases there are no visible differences between the 1$^{st}$ and 5$^{th}$ order models.\label{fig:larmor_power} 
}
\end{figure}

Having introduced the vector beam model, which provides an exact analytical solution to Maxwell's equations, we are now in a position to consider more extreme focussing, beyond the limit of validity of the paraxial approximation. Before considering emission spectra it makes sense to study more quantitatively  when the paraxial model starts to break down. In order to do this we need to choose a measurable quantity to enable comparisons between the models. An obvious choice is the total radiated power, given by the Larmor equation
\begin{equation}
P=\frac{2}{3} \frac{m r_e}{\gamma_0(1+\beta_0)} \int d\eta\, \ddot{x}^2, \label{eq:larmor}
\end{equation}
where $r_e=e^2/m$ is the classical electron radius and $\ddot{x}$ is the proper acceleration, distinct from the quantity $\dot{\beta}$ in Eq.~(\ref{eq:spectrum}). In Fig.~\ref{fig:larmor_power} we plot Eq.~(\ref{eq:larmor}) as a function of $\theta_0$ for various $a_0$ and $\gamma_0$.  Assuming that the vector beam model provides the ``correct'' solution (since it exactly satisfies Maxwell's equations) we find, as expected, that the paraxial model becomes less accurate as we increase $\theta_0$.  Both models exhibit the same downward trend in total radiated power as $\theta_0$ increases, which is hardly surprising because the pulse length becomes shorter. In all cases the paraxial model underestimates the radiated power, although the degree of underestimation depends on the parameters. We find the larger $\gamma$ is compared to $a_0$ the larger the error: for example, the case $a_0=2$, $\gamma_0=10$ (panel(c)) produces a relative error of approximately 25\% for $\theta_0=0.85$. The reason for this can be understood by considering the particle trajectory. We explained in the previous section that the transverse diameter of the electron orbit is proportional to $\gamma(1-\beta)$. Therefore the higher the $\gamma$ factor the narrower the orbit, meaning that the electron remains closer to the laser axis where the focussing effects (e.g. longitudinal electric field) are most significant.

\begin{figure}
\includegraphics[width=1.0\columnwidth]{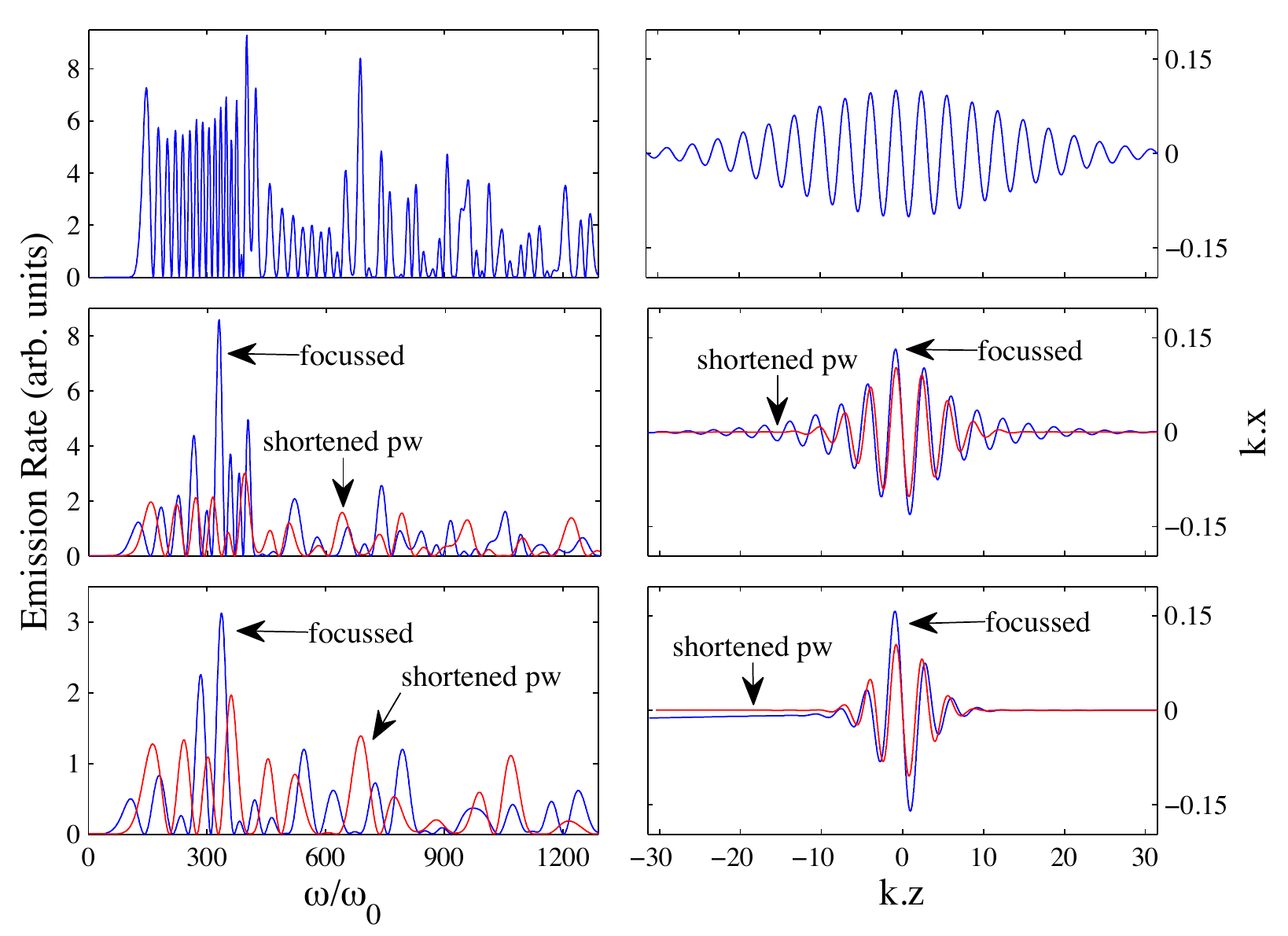}
\caption{Spectra and trajectories for an electron of initial $\gamma_0=10$ colliding with a laser of peak $a_0=2$, duration 10 cycles (27fs) and wavelength $\lambda=0.8\mu m$, for different levels of focussing. Left panels show emission spectra, right panels show the trajectories. Top row (blue lines): plane wave. Middle row (blue lines): vector beam focussed to $\theta_0=0.8$ ($w_0=0.318 \mu m$). Bottom row (blue lines): vector beam focussed to $\theta_0=1.2$ ($w_0=0.212 \mu m$). The red lines show the same but are calculated using a plane wave with duration shortened to match the {\it actual} FWHM duration of the focussed field (which becomes sortened by the focussing). For the middle panels this duration is 3.5 cycles, and for the bottom panels 2.5 cycles.\label{fig:tight_traj} }
\end{figure}

In Fig.~\ref{fig:tight_traj} we show the emission spectra and corresponding trajectories for electrons in lasers of increasingly tight focussing. (We model the tightly focussed pulses using the vector beam model.) It can be seen that, as with our earlier examples, the pulse duration becomes shorter as the focussing increases. This results in a lower amplitude for the emissions as well as structural changes to the harmonics. In order to distinguish between the effects of the pulse shortening and the other effects of the focussing, we measure the FWHM of the focussed pulse and calculate the spectra again using a plane wave of the same duration. The results for the corresponding plane wave fields are shown in gray. Comparing the focussed field results with the shortened plane wave results tells us that, while the shortening of the field goes a long way towards accounting for the reduction in amplitude and frequency range of the focussed pulse spectra, there are also structural changes in the harmonics caused by other focussing effects such as the longitudinal electric field and the curvature of the fields.

\begin{figure}
\includegraphics[width=1.0\columnwidth]{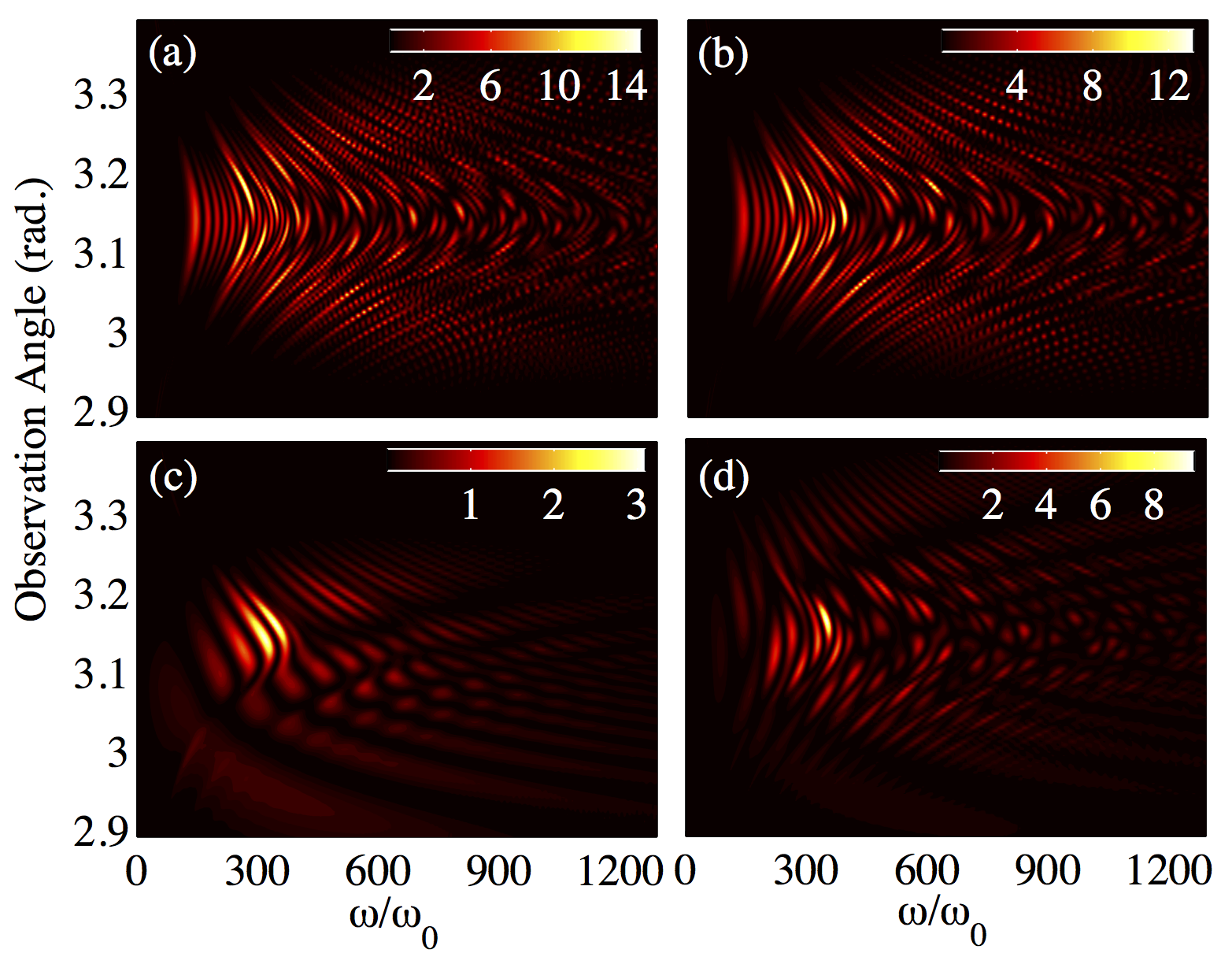}
\caption{Emission spectra for an electron of initial $\gamma_0=10$ in collision with a 10 cycle FWHM laser of peak $a_0=2$ and $\lambda=0.8 \mu m$. (a) Laser is a plane wave. (b) Laser is modelled as a 5th order paraxial beam focussed to a spot size of $w_0=0.8\mu m$ ($\theta_0=1/\pi$). (c) a 5th order paraxial beam focussed to a spot size of $w_0=0.2\mu m$ ($\theta_0=1.2$). (d) The same as (c) but calculated using the vector beam model.\label{fig:tight2d} }
\end{figure}

\begin{figure}
\includegraphics[width=1.0\columnwidth]{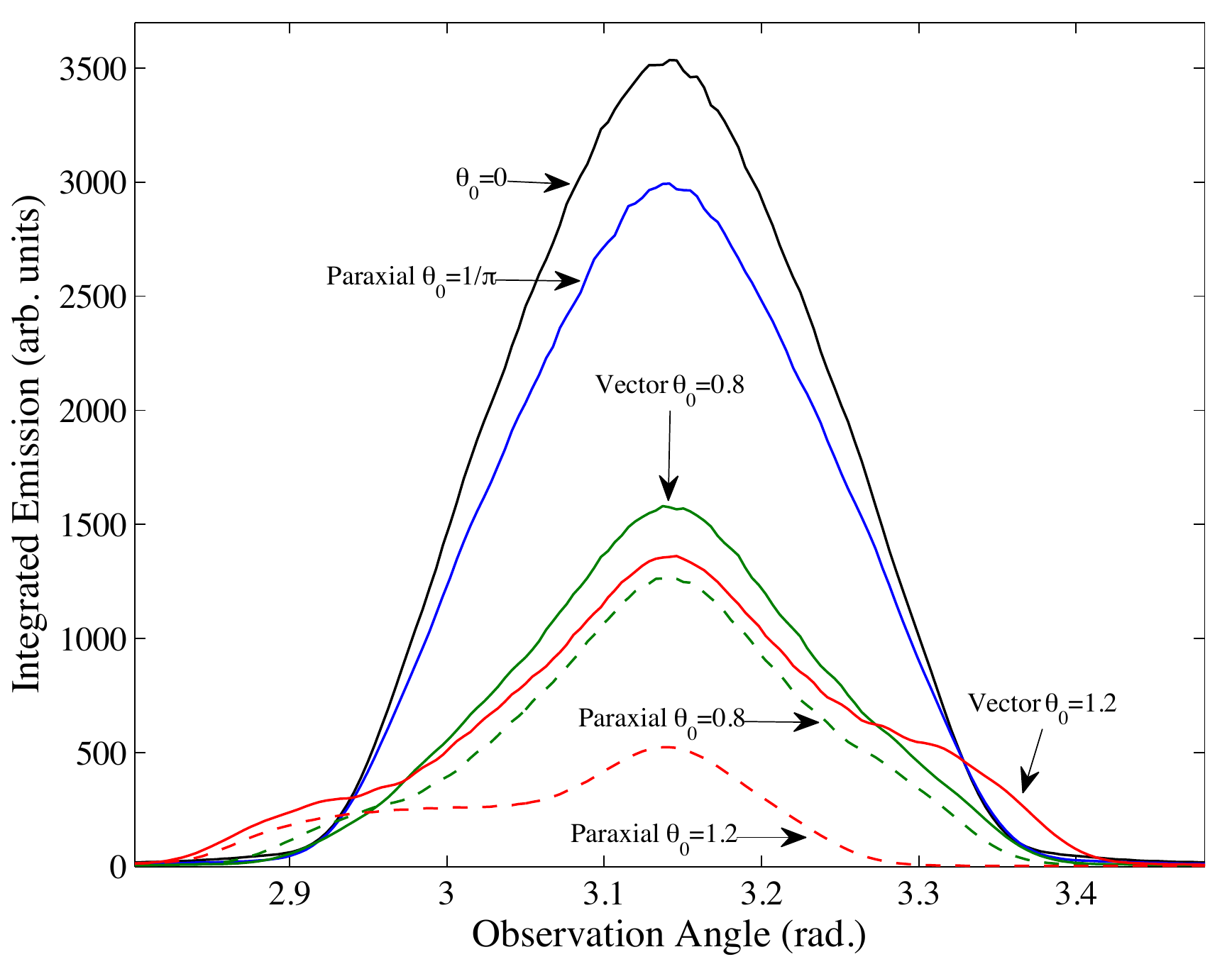}
\caption{Angular emission spectra integrated over all frequencies for fields of different focussing.  In all cases the laser has a peak $a_0=2$, $\lambda=0.8 \mu m$ and has an envelope function that is 10 cycles in duration. The incoming electron has an initial $\gamma_0=10$.  \label{fig:tight_theta} }
\end{figure}

In Fig.~\ref{fig:tight2d} we show the two-dimensional emission spectra as a function of both frequency and angle for different focussing parameters. It can be seen from the top panels that even reasonably strong focussing ($\theta_0=1/\pi$) results in only limited impact on the shape of the 2D spectral features. In the bottom panels we consider extremely tight focussing, $\theta_0=1.2$, implying that the beam is focussed to a spot of only half a wavelength in diameter. This is beyond the range of validity for which the paraxial expansion is valid, but for reference purposes we plot the spectrum for both the paraxial and vector beam models. It can be seen immediately that the (invalid) paraxial model produces much broader harmonic structures than the vector beam model. Nevertheless, both models predict an angular asymmetry about the laser axis (although this is exaggerated by the paraxial beam model). The reason for this asymmetry is that the duration of the pulse is so shortened by the focussing that the electron velocities becomes unsymmetrical in the peak of the pulse, just as we saw in the case of short pulses in Fig.~\ref{fig:ultra-short2D}.  Finally, in Fig.~\ref{fig:tight_theta} we show the integrated angular spectra for the four cases. It can be seen that the asymmetry starts to develop around $\theta_0=0.8$ and is over-estimated by the paraxial model. (For further discussion of the break down of the paraxial approximation we refer the reader to Ref.~\cite{Vaveliuk:2007}.)

%
%
\subsection{Effects at high intensities}\label{Sec:High-Intensity}

\begin{figure}
\includegraphics[width=1\columnwidth]{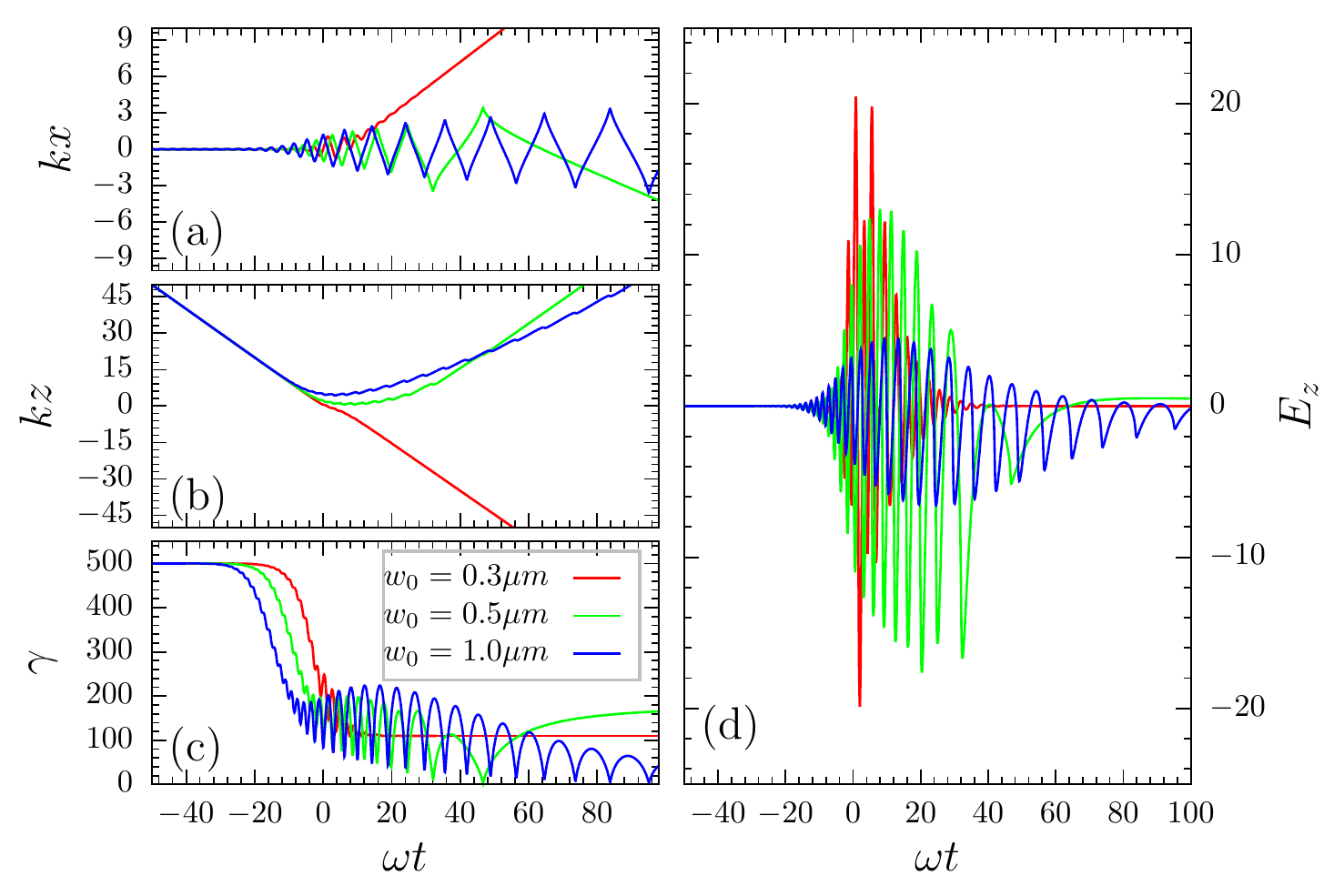}
\caption{Electron dynamics for three different focussing parameters, with RR included. These are $w_0=0.3\mu$m ($\theta_0=0.85$), $w_0=0.5\mu$m ($\theta_0=0.51$), and $w_0=1.0\mu$m ($\theta_0=0.25$). In all cases the laser is modelled as a paraxial beam of peak intensity $a_0 = 200$, wavelength $0.8\mu$m and duration 27fs. The electron has an initial $\gamma_0=500$ and is injected along the laser axis. The right hand panel shows the longitudinal electric field (normalized to $eE_z/\omega_0 m$) as experienced by the particle.
\label{fig:tra_high_intensity}}
\end{figure}
 
\begin{figure}
\includegraphics[width=1\columnwidth]{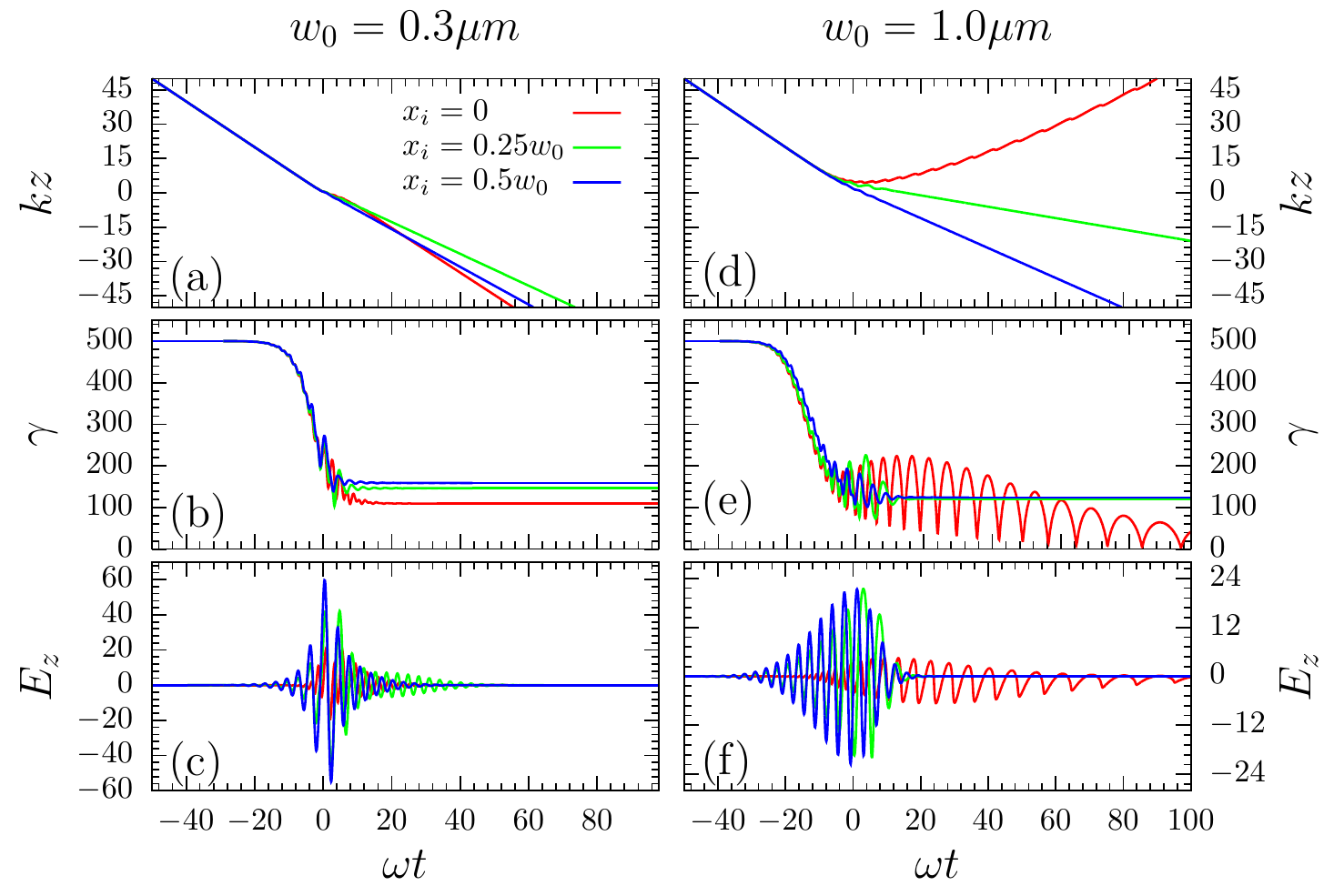}
\caption{Electron dynamics for three different impact parameters, with RR included. In all cases the laser is modelled as a paraxial beam of peak intensity $a_0 = 200$, wavelength $0.8\mu$m and duration 27fs. The electron has an initial $\gamma_0=500$ and is injected along the laser axis. The two lower panels show the longitudinal electric field (normalized to $eE_z/\omega_0 m$) as experienced by the particles.
\label{fig:tra_rr_xi}}
\end{figure}

In the cases we have considered up until now the total radiation emitted has been small enough for us to be able to neglect the resulting energy loss on the electron motion. However, if we consider more intense laser fields then the emissions will be of sufficiently high energy that the resulting energy losses to the particle will become significant.
This so-called `radiation reaction' effect can be included by adding correctional terms to the Lorentz force equation. 
However, determining the correct form of these terms is surprisingly non-trivial. Despite having been studied for over 100 years, it remains one of the most fundamental problems in electrodynamics.
A common starting point is to solve the coupled Lorentz and Maxwell's equations for the system. 
Doing so results in the Lorentz-Abraham-Dirac equation \cite{Lorentz:1905,Abraham:1905,Dirac:1938nz}, which is infamous due to its unphysical defects such as pre-acceleration and (unphysical) runaway solutions.  
One of the most common resolutions is to adopt a perturbative approximation, first proposed by Landau and Lifshitz \cite{LLII}.  Then the equation of motion is given by 
\begin{eqnarray}
\frac{d\mathbf{p}}{dt}=e(\mathbf{f}_\textrm{L}+\mathbf{f}_\textrm{R}),
\end{eqnarray}
where $\mathbf{f}_\textrm{L}=\mathbf{E}+\mathbf{v}\times\mathbf{B}$, and the radiative correction term is given by
\begin{eqnarray}
\mathbf{f}_\textrm{R}&=&-\bigg( \frac{4}{3} \pi \frac{r_e}{\lambda}  \bigg)
\bigg\{\gamma\left[ \left( \frac{\partial}{\partial t}+\mathbf{v}\cdot\nabla\right)\mathbf{E}+\mathbf{v}\times \left( \frac{\partial}{\partial t}+\mathbf{v}\cdot\nabla\right)\mathbf{B}\right]\nonumber\\
&&+ \bigg[(\mathbf{f}_L)\times\mathbf{B}+(\mathbf{v}\cdot\mathbf{E})\mathbf{E}
- \gamma^2 [(\mathbf{f}_L)^2-(\mathbf{v}\cdot\mathbf{E})^2]\mathbf{v}\bigg]\bigg\},\label{LL}
\end{eqnarray}
where $r_e=e^2/m$ is the classical electron radius.
Equation (\ref{LL}) is valid when the radiative reaction force is much less than the Lorentz force in the instantaneous rest frame of the particle.  We note that there are a growing number of alternative equations in the literature (for an overview see \cite{Burton:2014,Vranic:2015}) and it is still an open problem as to which is the correct formulation.  However, all of the models predict almost indistinguishable particle dynamics \cite{PhysRevE.88.011201}. Additionally, it has recently been shown that the Landau-Lifshitz equation, along with some of the others, is consistent with quantum electrodynamics to the order of the fine-structure constant $\alpha$ \cite{0038-5670-34-3-A04,Ilderton:2013tb}.
Finally, we note that the first term (derivative term) of Eq.~(\ref{LL}) is significantly smaller than the other two, since it is only linear in the field strength whereas the other terms are quadratic.  We find that in all cases the contribution from this term is negligible and so we do not include it in our simulations. (In fact, it can be shown that, in the cases where classical RR is important, the derivative term is even smaller than the electron spin force and so one can argue that it should be neglected out of consistency~\cite{Tamburini:2010}.)

In Fig.~\ref{fig:tra_high_intensity} we plot various aspects of the electron dynamics for different levels of pulse focussing. In all the cases we take a much stronger laser pulse than before, with a peak intensity of $a_0=200$ (corresponding to $1.2\times10^{23}$W/cm$^2$). The electron has an initial $\gamma_0=500$ and is injected along the laser axis, providing parameters such that RR effects are significant. This is evident from Fig.~\ref{fig:tra_high_intensity}(c) where it can be seen that the electrons lose almost all of their initial energy as they travel through the laser pulse. In Fig.~\ref{fig:tra_high_intensity}(b) we see that for weaker focussing the electrons lose so much energy that they become reflected by the laser pulse \cite{Harvey:gaussian:2012}. In the case of very tight focussing this is no longer the case, the reason being once again that the pulse length is significantly shortened by the focussing and so the electrons are decelerated for a shorter period of time. Although it's true that the more tightly focussed pulse has a stronger longitudinal electric field, this is not enough to compensate for the shortened pulse duration. In Fig.~\ref{fig:tra_high_intensity}(d) we demonstrate this by plotting the (normalised) longitudinal field as seen by the electron. It can be seen that, as expected, the tighter the focussing the stronger the peak $E_z$ field. However, the weaker the focussing the longer the time the electron is exposed to this field  which means that, somewhat counterintuitively, the impulse imparted to the electron from the longitudinal field is greater (this is particularly true in the case where the electron is reflected and starts co-propagating with the laser pulse). We also note that for the most tightly focussed pulse the electron is given a strong transverse kick by the ponderomotive effect resulting from the strong field gradients.

\begin{figure}
\begin{center}
\includegraphics[width=1\columnwidth]{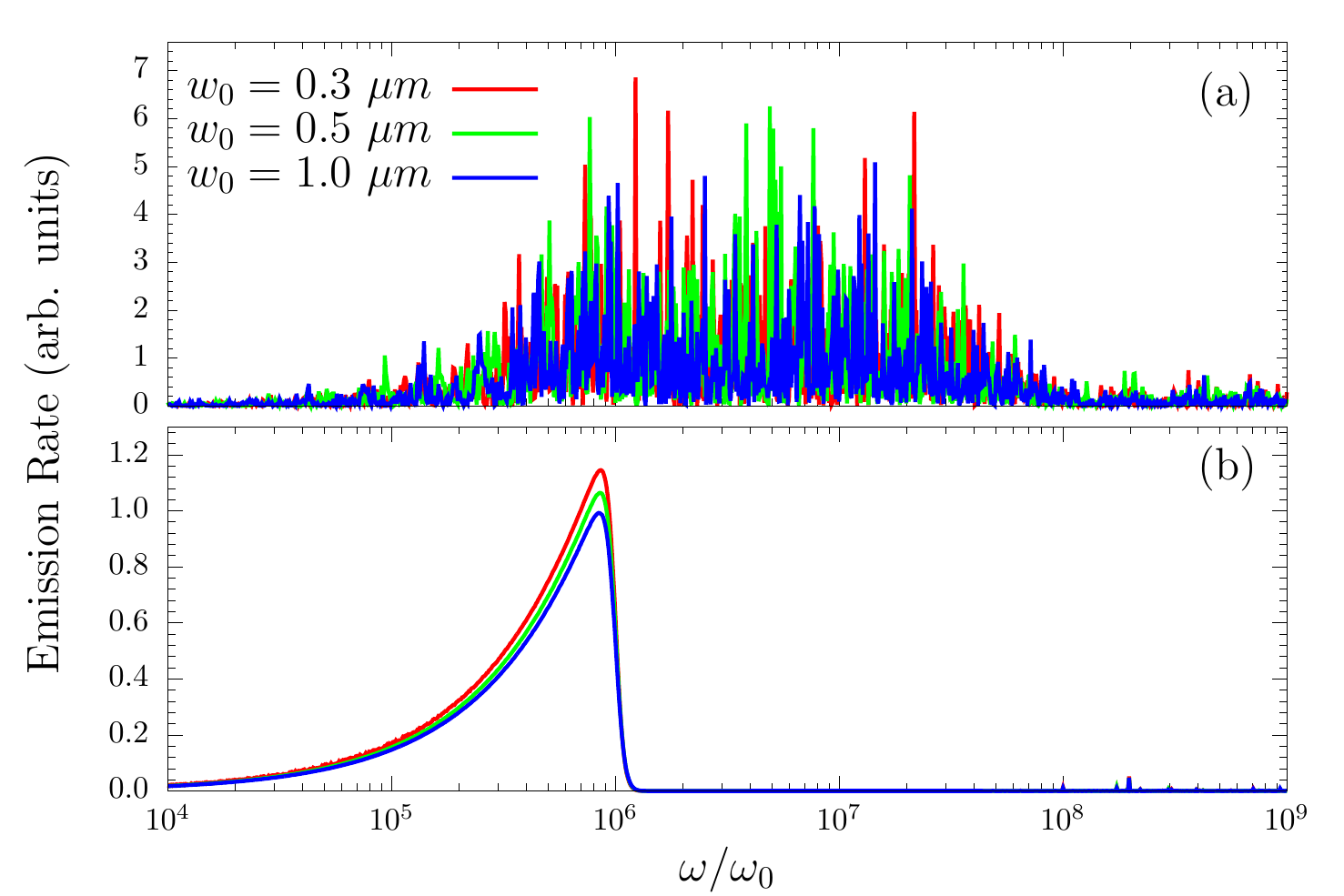}
\caption{Thomson emission spectra calculated in the backscattering direction, $\theta_{\textrm{obs}}=\pi$, for the electrons in Fig.~\ref{fig:tra_high_intensity}. Panel (a) shows the spectra for the exact setup described in Fig.~\ref{fig:tra_high_intensity}, while panel (b) shows the spectra calculated using a circularly polarised laser pulse with the same peak intensity and duration.\label{fig:spec_rr_lp_cp}}
\end{center}
\end{figure}

Before turning our attention to the emission spectra we also briefly consider the effects of impact parameter on the motion of an electron in a pulse where RR is important. Taking the same parameters as Fig.~\ref{fig:tra_high_intensity}, in Fig.~\ref{fig:tra_rr_xi} we plot dynamical quantities for electrons injected along the laser axis and offset by a quarter and a half of the waist radius. We find (left column) that in the case of strong focussing there is little variation in the particle motion as we change the impact parameter. This is because the shortening of the pulse due to the focussing means that the electrons don't have time to lose enough energy to be strongly affected by the structure of the laser pulse. Conversely in the more weakly focussed pulse (right column) the electron that passes right through the centre is in the field long enough to be reflected, whereas the particles with non-zero impact parameters see a weaker pulse (see Sec.~\ref{Sec:impact}) and continue more or less in their original direction of motion.

Having considered the electron motion we are now ready to study the resulting Thomson emission spectra. In Fig.~\ref{fig:spec_rr_lp_cp}(a) we show the emission spectra calculated in the backscattering direction, $\theta_{\textrm{obs}}=\pi$, for the electrons in Fig.~\ref{fig:tra_high_intensity}. While the motion of the electrons was found to be strongly affected by the level of focussing, we see that there is very little difference between the emission spectra. In all cases they rise in amplitude at around $10^5\omega_0$ and die off around $10^8\omega_0$. The reason for this discrepancy is that most of the radiation emissions occur {\it while} the electrons are slowing down; it is only after they have lost energy to emissions that they become susceptible to being reflected (or deflected by the pulse structure). Thus the rich dynamics we see in the electron motion as a result of the pulse focussing have little effect on the Thomson radiation spectra. We note that there are specific cases where the combination of focussing and RR becomes important, such as for the attosecond $\gamma$ ray source described in Ref.~\cite{Li:2015atto}, but in general the effect is minimal. We further illustrate this in Fig.~\ref{fig:spec_rr_lp_cp}(b) where we show the same again, but this time for the case of a circularly polarised laser. In the case of circular polarisation only the first harmonic contributes to the spectrum in the direction $\theta_{\textrm{obs}}=\pi$ \cite{Harvey:2009}, whereas in the case of linear polarisation all of the odd numbered harmonics contribute \cite{Esarey:1993}. The result is that the spectra from the circularly polarised field is much cleaner, of shorter frequency range, and easier to analyse. (This is also the reason why the chirped Thomson spectra evaluated at $\theta_{\textrm{obs}}=\pi$ in Ref.~\cite{Holkundkar:2015} qualitatively differ from the integrated spectra in Ref.~\cite{Yoffe:2016}.) We can see that in all three cases the spectra die off at the same frequency and, apart from a modest change in amplitude, there is little dependence on the laser focussing.

\section{Conclusions}\label{Sec:Conclusions}
In this paper we have assessed the effects of laser pulse focussing on the spectral properties of Thomson scattered radiation. By comparing the spectra obtained using a paraxial field with a plane wave model we found that, in all but the most extreme focussing, the temporal envelope has a much bigger effect on the spectrum than the focussing itself. This is relevant to the modelling community since it provides assurance that plane wave models can be used to obtain a reasonable understanding of the properties of the spectra. This is particularly important in studies involving strong field quantum electrodynamics where including focussing effects is still a largely unsolved problem.
Extending our analysis to the case of a spatially distributed bunch of electrons colliding with a focussed laser pulse, we found that a reasonable approximation can be found by replacing the field with a transversally decaying plane wave. The emission spectrum from a single electron on-axis also provides a good indication of the location of the first harmonic, but it contains higher harmonic structures which are not representative of the emissions from the rest of the electron bunch. This is because the off axis electrons pass through a lower peak field, and it means that the the higher frequency part of the spectrum is not well approximated by a single electron.

Next we considered cases where the pulse duration is very short. In such situations the paraxial model is no longer valid because it is derived assuming that the fields vary slowly along the propagation direction. Instead we adopt a sub-cycle vector beam model which describes a sub-cycle focussed field and exactly solves Maxwell's equations. (When the pulse duration in this description is increased to several cycles one re-obtains the paraxial solution.) Using this model we find that, as expected, the paraxial model becomes unreliable for pulses of duration less than one cycle. As the pulse duration decreases the emission harmonics become blue shifted and broaden out in frequency space. We interpret this as being a result of the lack of periodicity, meaning that the intensity dependent mass shift no longer plays a role, a finding consistent with \cite{PhysRevA.83.032106}. Additionally for very short pulses the carrier envelope phase becomes important, resulting in an angular asymmetry in the spectrum.

Armed with the vector beam model that exactly solves Maxwell's equations, we studied the effects of focussing beyond the limit at which the paraxial approximation breaks down. By comparing the total radiated power (Larmor power) we find that the two models begin to diverge when the focussing parameter $\theta_0\gtrsim 0.3$. (However, we note that this figure has been found to be even lower in the co-propagating case where the field structure is more important \cite{Salamin:2002}.)  As we increased the focussing to spot sizes that are below a wavelength in diameter we found that focussing causes the fields to die off so quickly that they behave qualitatively similar to sub-cycle fields. In particular we found once again that the carrier envelope phase becomes important and can cause an angular asymmetry in the emission spectra.

Finally, we turned our attention to high-intensity fields where radiation reaction effects become important to the particle motion. We found that, although the longitudinal electric fields that are present in the focussed pulse cause the electron to be reflected earlier than would otherwise be the case from radiation reaction alone, the focussing itself has limited impact on the emission spectrum. This is because most of the radiation is emitted before the particle loses energy, and it is only once this has happened that the radiation reaction  and focussing play a big role in the particle dynamics.

\section*{Acknowledgements}
CH and MM acknowledge support from the Swedish Research Council, grants 2012-5644 and 2013-4248. AH acknowledges the Science and Engineering Research Board, Department of Science and Technology, Government of India for funding the project SR/FTP/PS-189/2012.  
%
\appendix
\section{Modelling the laser field}\label{Sec:Fields}

\subsection{Paraxial Approximation}\label{Sec:Paraxial}
We begin with the most commonly used description of a focussed laser pulse, the paraxial Gaussian beam.  This describes a focussed electromagnetic field evolving in time and space, which satisfies Maxwell's equations to the order of an expansion parameter proportional to the ratio of the field wavelength to the beam waist. We now briefly summarise the field's derivation to fifth order (for further details see Refs.~\cite{Salamin:2002}).
Throughout this article we work in natural units where $\hbar=c=1$.
We work in the Lorentz gauge so that the vector potential of the field $A=(\phi,\mathbf{A})$ satisfies
\begin{eqnarray}
\frac{\partial\phi}{\partial t}+\boldsymbol{\nabla}\cdot\boldsymbol{A}=0.\label{lorentzgauge}
\end{eqnarray}
Additionally the vector potential must satisfy the vacuum wave equation
\begin{eqnarray}
\nabla^2\boldsymbol{A}=\frac{\partial^2\boldsymbol{A}}{\partial t^2}. \label{vacuumwaveeqn}
\end{eqnarray}
Taking the laser to be linearly polarised in $x$ and propagating in the $+z$ direction, we write the potential as 
\begin{eqnarray}
\boldsymbol{A}=\boldsymbol{\hat{x}}A_0 a(\eta)\psi (x,y,z)e^{-ikz},\label{A1}
\end{eqnarray}
where $A_0$ is the amplitude of the pulse, $\eta =\omega t-kz$, and $a(\eta)$ is a generic pulse shape function.  Inserting (\ref{A1}) into (\ref{vacuumwaveeqn}) gives us
\begin{eqnarray}
\nabla^2\psi-2ik\frac{\partial\psi}{\partial z} 
\left ( 1-i\frac{a^\prime}{a}\right )=0, \label{psieqn}
\end{eqnarray}
where $a^\prime =da/d\eta$.  In general it is hard to satisfy (\ref{psieqn}) since $\psi$ is a function of all three spatial coordinates $(x,y,z)$.  To proceed we renormalise the coordinates
\begin{eqnarray}
\xi\equiv\frac{x}{w_0}, \quad \nu\equiv\frac{y}{w_0}, \quad \zeta\equiv\frac{z}{z_r},
\end{eqnarray}
making them dimensionless.  In doing so we have introduced the beam waist diameter $w_0$ and the Rayleigh length $z_r=kw_0^2/2$.  Imposing one further constraint, that the pulse shape function satisfies
\begin{eqnarray}
a^\prime \ll a, \label{gconstraint}
\end{eqnarray}
equation (\ref{psieqn}) can be approximated by
\begin{eqnarray}
\nabla_\perp^2\psi -4i\frac{\partial\psi}{\partial\zeta}+\theta_0^2
\frac{\partial^2\psi}{\partial\zeta^2}=0, \label{psieqnapprox}
\end{eqnarray}
where
\begin{eqnarray}
\nabla_\perp^2 =\frac{\partial^2}{\partial\xi^2}+\frac{\partial^2}{\partial\nu^2},\quad
\psi=\psi(\xi, \nu, \zeta ),
\end{eqnarray}
and we have introduced the aspect ratio $\theta_0 =w_0/z_r=\lambda/\pi w_0$ which, when small, closely approximates the beam diffraction angle.  Assuming that $\theta_0$ is small, or in other words that the focussing is not too strong, we can expand $\psi$ in the series
\begin{eqnarray}
\psi=\psi_0+\theta_0^2\psi_2+\theta_0^4\psi_4+\ldots.
\end{eqnarray}
Equating coefficients of $\theta_0$ we have, from (\ref{psieqnapprox}),
\begin{eqnarray}
\nabla_\perp^2\psi_0-4i\frac{\partial \psi_0}{\partial\zeta} &=& 0, \label{E1}\\
\nabla_\perp^2\psi_2-4i\frac{\partial \psi_2}{\partial\zeta}+\frac{\partial^2 \psi_0}{\partial\zeta^2} &=& 0,\label{E2}\\
\nabla_\perp^2\psi_4-4i\frac{\partial \psi_4}{\partial\zeta}+\frac{\partial^2 \psi_2}{\partial\zeta^2} &=& 0,\label{E3}\\
\textrm{etc.}\nonumber
\end{eqnarray}
Equation (\ref{E1}) is the well-known paraxial wave equation with solution
\begin{eqnarray}
\psi_0 =be^{-b\rho^2},
\end{eqnarray}
where
\begin{eqnarray}
b=\frac{1}{\sqrt{1+\zeta^2}}e^{i\arctan\zeta},\quad \rho^2=\xi^2+\nu^2.
\end{eqnarray}
The solution to (\ref{E2}) was originally found by Davis \cite{Davis:1979}
\begin{eqnarray}
\psi_2=\left ( \frac{b}{2}+\frac{b^3\rho^4}{4}\right )\psi_0,
\end{eqnarray}
and Barton and Alexander \cite{Barton:1989} proceeded to find the solution to (\ref{E3})
\begin{eqnarray}
\psi_4=\frac{1}{32}(12b^2-6b^4\rho^4-4b^5\rho^6+b^6\rho^8)\psi_0.
\end{eqnarray}
Before we can calculate the field components we also need to know the scalar potential. Just as with the vector potential (\ref{A1}) we start by assuming that this can be written in the form
\begin{eqnarray}
\phi(t,x,y,z)=a(\eta)\Phi(x,y,z)e^{i\eta}.
\end{eqnarray}
The Lorentz gauge condition (\ref{lorentzgauge}) then gives us
\begin{eqnarray}
\frac{\partial\phi}{\partial t}=i\omega\phi\left (1-i\frac{a^\prime}{a}\right )\approx i\omega\phi,
\end{eqnarray}
which means that
\begin{eqnarray}
\phi=\frac{i}{k}\boldsymbol{\nabla}\cdot\boldsymbol{A}.
\end{eqnarray}
Now the electric and magnetic field components can be calculated from (\ref{A1}) via
\begin{eqnarray}
\boldsymbol{E} &=& -ik\boldsymbol{A}-\frac{i}{k}\boldsymbol{\nabla}(\boldsymbol{\nabla}\cdot\boldsymbol{A}),\\
\boldsymbol{B} &=& \boldsymbol{\nabla}\times\boldsymbol{A},
\end{eqnarray}
(for details of the calculation see \cite{Salamin:2002, Barton:1989}).  Taking the real part of the resulting expressions gives us (to fifth order in $\theta_0$)
\begin{align}
E_x =& P\big(S_0 + \frac{\theta_0^2}{4}\big [4\xi^2 S_2 -\rho^4S_3\big ]+\frac{\theta_0^4}{32}\big[  4S_2-8\rho^2 S_3\nonumber\\ 
&-2\rho^2(\rho^2-16\xi^2)S_4 -4\rho^4(\rho^2+2\xi^2)S_5 +\rho^8S_6\big ]\big),\label{fieldEx}\\
E_y =& P\xi\nu\big(\theta_0^2 S_2+\frac{\theta_0^4}{4}\big [ 4\rho^2 S_4-\rho^4 S_5 \big ]         \big),\label{fieldEy}\\
E_z =& P\xi\big( \theta_0 C_1  +\frac{\theta_0^3}{4}\big [ -2C_2+4\rho^2 C_3-\rho^4 C_4 \big ] \nonumber \\
&+\frac{\theta_0^5}{32} \big [ -12C_3 -12\rho^2 C_4 +34\rho^4 C_5 \nonumber \\
&-12\rho^6 C_6 +\rho^8 C_7   \big ]  \big),\label{fieldEz}\\
B_x =& 0,\label{fieldBx}\\
B_y =& P\big( S_0+ \frac{\theta_0^2}{4}\big [  2\rho^2 S_2-\rho^4S_3\big ]
+\frac{\theta_0^4}{32}\big [  - 4S_2+8\rho^2S_3\nonumber\\ 
&+10\rho^4 S_4-8\rho^6 S_5 +\rho^8 S_6\big ]   \big),\label{fieldBy}\\
B_z =& P\nu\big( \theta_0 C_1 +\frac{\theta_0^3}{4}\big [ 2C_2+2\rho^2 C_3-\rho^4 C_4  \big ]  \nonumber\\ 
&+\frac{\theta_0^5}{32}\big [  12C_3 +12\rho^2 C_4 +6\rho^4C_5 -8\rho^6 C_6 +\rho^8C_7   \big ]\big),\label{fieldBz}
\end{align}
where the prefactor is given by
\begin{eqnarray}
P=A_0 \frac{w_0}{w}a(\eta )\textrm{exp}\big( -\frac{r^2}{w^2} \big), \quad r^2=x^2+y^2. \label{eq:P}
\end{eqnarray}
Here $w=w(z)$ is a measure of the beam diameter at a given longitudinal coordinate
\begin{eqnarray}
w(z)=w_0\sqrt{1+\bigg(\frac{z}{z_r}\bigg)^2}.
\end{eqnarray}
Finally, the functions $S_j$ and $C_j$ are defined
\begin{eqnarray}
S_j &=& \bigg( \frac{w_0}{w} \bigg)^j \sin\Theta, \\
C_j &=& \bigg( \frac{w_0}{w} \bigg)^j \cos\Theta, 
\end{eqnarray}
where
\begin{eqnarray}
\Theta =\eta-\frac{kr^2}{2H}+(j+1)\arctan\zeta,
\end{eqnarray}
where $H=z+z_r^2/z$ is the radius of curvature of the field.

The electric and magnetic field components (\ref{fieldEx}-\ref{fieldBz}) describe the laser to fifth order in $\theta_0$.  For an optical laser of wavelength $\lambda=0.8\mu$m focussed to a waist size $w_0=5\mu$m the expansion parameter is
\begin{eqnarray}
\theta_0=\frac{\lambda}{\pi w_0}=\frac{0.8}{5\pi}\approx 0.05.
\end{eqnarray}

\subsection{Plane Wave Limit}

In the limit where the beam waist becomes large, $w_0\to\infty$, we obtain the plane wave expressions 
\begin{eqnarray}
E_x &=& A_0 a(\eta)\sin\eta,\quad E_y =0,\quad E_z=0, \label{Eplanewave}\\
B_x &=&  0, \quad B_y =A_0 a(\eta)\sin\eta,\quad B_z=0.\label{Bplanewave}
\end{eqnarray}
Such fields are infinite in their transverse spatial extent. They also can be considered to be perfectly polarized, exhibiting no field components in either the longitudinal or unpolarized transverse direction. This contrasts with the focussed beam which has additional field components in both these directions, with amplitudes proportional to the focussing parameter $\theta_0$.

\subsection{Focussed Vector Beams}\label{Sec:Vector_Beams}
The paraxial beam expansion is limited in its range of validity. 
In Sec.~\ref{Sec:Paraxial} we consider terms to fifth order in $\theta_0$. Although higher-order terms have been derived \cite{Salamin:2006}, in order to go to tighter focussing a different approach must be taken.
If the beam is focussed too strongly then the expansion parameter $\theta_0$ will no longer be able to be considered small, calling into question the convergence of the series. 
Additionally, in the case where the pulse duration is very short the paraxial approximation will no longer be valid. This is because in such a situation the expansion parameter $\theta_0$ (which closely approximates the beam diffraction angle) will be of a similar order to the the timescale of the pulse duration $\omega_0 T$, and so the fields can no longer be considered to vary gradually along the propagation axis.
An alternative approach is that of the focussed vector beam model derived in Ref.~\cite{Lin:2006}.

The vector beam model is derived via the use of the complex source method \cite{Heyman:1989,Deschamps:1971,Wang:2003} and provides an exact analytical solution in closed form satisfying Maxwell's equations. In the case of linear polarization, one begins by considering an oscillating dipole at the coordinate origin with dipole moment
\begin{equation}
\mathbf{p}(\mathbf{r},t)=p_0(t)\hat{x}\delta(\mathbf{r}).
\end{equation}
The function $p_0(t)$ can be defined arbitrarily, but for our purposes it is taken to be a oscillating wave with a Gaussian carrier envelope of FWHM duration $2\sqrt{2 \ln (2)} T$.
\begin{equation}
p_0(t) = p_0 \exp\bigg( -\frac{ t^2}{2T^2}\bigg) \exp (i\omega t+ i\phi_0),
\end{equation}
where $p_0=4\pi z_r M_0 E_0/k^2$ is the peak power of the beam, and $M_0=\sqrt{N/(N+1/k^2z_r^2)}$, where $N=1-1/kz_r+1/\omega^2T^2$.
This dipole emits a spherical electromagnetic pulse. In order to obtain a propagating focussed pulse the authors of \cite{Lin:2006} introduce a complex coordinate shift to $z$ and $t$
\begin{align}
z& \rightarrow z +iz_r,\\
t& \rightarrow t-t_0+i z_r. 
\end{align}
The result is a moving field structure which can be used to derive the expressions for a propagating pulse. Upon performing the algebra, the resulting field components are found to be
\begin{align} 
E_x =& \textrm{Re} \bigg[M \bigg( f+\frac{x^2g}{R^2} \bigg) \bigg ], \\ 
E_y =& \textrm{Re} \bigg[M \bigg( \frac{xy g}{R^2} \bigg) \bigg ], \\ 
E_z =& \textrm{Re} \bigg[M  f+\frac{xz g}{R^2}  \bigg ], \\ 
B_x =& 0, \\
B_y =& \textrm{Re} \bigg[M  \frac{z }{R }h  \bigg ], \\   
B_z =& \textrm{Re} \bigg[-M  \frac{y}{R }h  \bigg ],
\end{align}
where 
\begin{align}
f =& \bigg( 1 + \frac{i\eta^\prime}{\omega T^2}\bigg)^2-\frac{1}{k^2 R^2}\bigg( 1-\frac{t^\prime R}{T^2}+ikR\bigg),\\
g=& -f + \frac{2}{k^2R^2}\bigg( 1-\frac{\eta^\prime R}{T^2}+ikR\bigg),\\
h=& f + \frac{1}{k^2R^2},
\end{align}
the quantity $R=\sqrt{x^2+y^2+(z+iz_r)^2}$ expresses a complex distance and the complex retarded phase is given by $\eta^\prime=t-R$. Finally, the prefactor  $M=E_0 z_r M_0 p_0 (\eta^\prime )/p_0 R$.
It can be easily shown \cite{Lin:2006} that in the limit where the beam waist $w_0 \rightarrow\infty$ then one recovers the plane wave fields (\ref{Eplanewave}-\ref{Bplanewave}). Additionally, if one expands the fields in $\theta_0$ one finds that the terms agree with those of the paraxial model (\ref{fieldEx}-\ref{fieldBz}).

%
%

\end{document}